\documentclass[a4paper]{article}%
\usepackage{amsmath}
\usepackage{amssymb}
\usepackage{amsfonts}
\usepackage[T1]{fontenc}
\usepackage{graphicx}%
\usepackage{epsfig}%
\usepackage{calc}%
\usepackage{dsfont}
\usepackage{longtable}%
\usepackage{overcite}%
\usepackage{lscape}%
\usepackage{multirow}%

\newcommand{\pa}[1]{\left( #1 \right)}
\newcommand{\br}[1]{\left[ #1 \right]}
\newcommand{\ac}[1]{\left\{ #1 \right\}}

\newcommand{\al}{\alpha}
\newcommand{\bb}{\beta}

\newcommand{\dd}{\delta}

\newcommand{\ve}{\varepsilon}

\newcommand{\vf}{\varphi}

\newcommand{\g}{\gamma}

\newcommand{\la}{\lambda}

\newcommand{\s}{\sigma}

\newcommand{\q}{\theta}

\newcommand{\w}{\omega}
\newcommand{\W}{\Omega}

\newcommand{\RR}{\mathds{R}}

\newcommand{\rank}{\operatorname{rank}}

\newcommand{\ie}{\textit{i.e.}\ }

\newcommand{\del}{\partial}

\newenvironment{aleq}{\begin{equation}\begin{aligned}}{\end{aligned}\end{equation}}
\newenvironment{aleq*}{\begin{equation*}\begin{aligned}}{\end{aligned}\end{equation*}}
\newenvironment{gaeq}{\begin{equation}\begin{gathered}}{\end{gathered}\end{equation}}
\newenvironment{gaeq*}{\begin{equation*}\begin{gathered}}{\end{gathered}\end{equation*}}
\newenvironment{eqe}{\begin{equation}}{\end{equation}}

\makeatletter
\renewcommand{\@biblabel}[1]{$^{#1}$}%
\makeatother%
\begin{document}%
\title{\bf Symmetry group analysis of an ideal plastic flow}
\author{Vincent Lamothe\thanks{email address: lamothe@crm.umontreal.ca}\\
D\'epartement de math\'ematiques et statistiques, Universit\'e de Montr\'eal,\\
C.P. 6128, Succc. Centre-ville, Montr\'eal, (QC) H3C 3J7, Canada}%
\date{}
\maketitle%
\fontsize{12}{14} \selectfont
\begin{abstract}In this paper, we study the Lie point symmetry group of a system describing an ideal
plastic plane flow in two dimensions in order to find analytical solutions. The infinitesimal
generators that span the Lie algebra for this system are obtained. We completely classify the
subalgebras of up to codimension two in conjugacy classes under the action of the symmetry group.
Based on invariant forms, we use Ansatzes to compute symmetry reductions in such a way that the
obtained solutions cover simultaneously many invariant and partially invariant solutions. We
calculate solutions of the algebraic, trigonometric, inverse trigonometric and elliptic type. Some
solutions depending on one or two arbitrary functions of one variable have also been found. In some
cases, the shape of a potentially feasible extrusion die corresponding to the solution is deduced.
These tools could be used to thin, curve, undulate or shape a ring in an ideal plastic material.
\end{abstract}
{Running Title: Symmetry group analysis of an ideal plastic flow\\
PACS numbers: Primary 62.20.fq; Secondary 02.30.Jr\\
Keywords: symmetry group of partial differential equations, symmetry reduction, invariant
solutions, ideal plasticity, extrusion die}
\section{Introduction}
In this paper, we investigate the plane flow of ideal plastic materials \cite{Kat:1,Hill,Chak}
modelled by the hyperbolic system of four partial differential equations (PDE) in $q=4$ dependent
variables $\s, \theta, u,v$ and $p=2$ independent variables $x$ and $y$,
\begin{aleq}\label{eq:1}
&(a)\qquad &&\s_x-2 k \pa{\q_x \cos 2\q+\q_y \sin2\q}=0,\\
&(b)\qquad &&\s_y-2 k \pa{\q_x\sin2\q - \q_y \cos 2\q}=0,\\
&(c)\qquad &&(u_y+v_x)\sin 2\q + (u_x-v_y)\cos2\q=0,\\
&(d)\qquad &&u_x+v_y=0,
\end{aleq}%
where $\s_x=\del \s/\del x$, \textit{etc}. The expressions (\ref{eq:1}.a), (\ref{eq:1}.b), are the
equilibrium equations for the plane problem. In other words they are the Cauchy differential
equations of motion in a continuous medium where we consider that the sought quantities do not
depend on $z$. These two equations involve the dependent variables $\s$ and $\q$ that define the
stress tensor; $\s$ is the mean pressure and $\q$ is the angle relative to the $x$ axis in the
counterclockwise direction minus $\pi/4$. The equation (\ref{eq:1}.c) corresponds, in the plane
case, to the Saint-Venant-Von Mises plasticity theory equations, where $u$ and $v$ are respectively
the velocities in the $x$ axis and $y$ axis directions. Moreover, we assume incompressibility of
the material and hence the velocity vector is divergenceless. This explains the presence of the
equation (\ref{eq:1}.d) in the considered system. The positive-definite constant $k$ is named the
volumetric compression coefficient and is related to the Poisson coefficient and the Young modulus
$\nu$ by the formula $k=(1-2\nu)E^{-1}$.
\paragraph*{}In order to calculate new solutions of the system
consisting of (\ref{eq:1}.a) and (\ref{eq:1}.b), S.I. Senashov \textit{et
al.}\cite{Senashov:2007,Senashov:2009} acted with transformations of the symmetry group of this
system on known solutions of some boundary value problems, \ie the Nadaï solution
\cite{Nadai:circularSol} for a circular cavity under normal stress and shear and the Prandtl
solution \cite{Prandtl:solPlas} for a bloc compressed between two plates. In addition, Czyz
\cite{Czyz:1} found simple and double wave solutions for the system (\ref{eq:1}) using the method
of characteristics. However, as it is often the case with this method, his solutions rely on
numerical integration for obtaining the velocities $u$ and $v$. To our knowledge, no systematic Lie
group analysis based on a complete subalgebra classification in conjugacy classes under the action
of the symmetry group $G$ of the system (\ref{eq:1}) has been done before.
\paragraph*{}The goal of this paper is to systematically investigate the system (\ref{eq:1}) from the perspective of the Lie group of point
symmetries $G$ in order to obtain analytical solutions. That is, we obtain in a systematic way all
invariant and partially invariant (of structure defect $\delta=1$ in the sense defined by
Ovsiannikov \cite{Ovsiannikov:Group_Analysis}) solutions under the action of $G$ which are
non-equivalent. Invariant solutions are said non-equivalent if they cannot be obtained one from
another by a transformation of $G$ (the solutions are not in the same orbit). In practice, we apply
a procedure developed by J. Patera \textit{et al}. \cite{PateraWinter:1,PateraWinter:2,WinterPIEEP}
that consists of classifying the subalgebras of $\mathcal{L}$ associated with $G$ into conjugacy
classes under the action of $G$. Two subalgebras $\mathcal{L}_i\subset \mathcal{L}$ and
$\mathcal{L}_i'\subset \mathcal{L}$ are conjugate if $G\mathcal{L}_iG=\mathcal{L}_i'$. For each
conjugacy class, we choose a representative subalgebra, find its invariants and use them to reduce
the initial system (\ref{eq:1}) to a system in terms of the invariants which involve fewer
variables. According to the approach proposed by Kruskal and Clarkson \cite{Clarkson_Kruskal},
which is part of the more general framework of conditional symmetries, we propose in this paper
some Ansatzes which allow us to cover simultaneously many invariant and partially invariant
solutions (PIS). These more general solutions reduce to invariant and partially invariant ones for
appropriate parameter values. We illustrate these theoretical considerations with many classes of
solutions. A more exhaustive collection of such solutions is provided in \cite{vincentPhDTh}.
Thereafter, we draw for some solutions the shape of the corresponding extrusion die. The applied
method relies on the fact that the walls of the tools must coincide with the flow lines describe by
the velocities $u$ and $v$ of the solutions of the problem. In application, it is convenient to
feed in material the extrusion die rectilinearly at constant speed. So, the tools illustrated in
this paper were drawn considering this kind of feeding. Based on mass conservation and on the
incompressibility of the materials, we easily deduce that the curve defining the limit of the
plasticity region for constant feeding speed must obey the ordinary differential equation (ODE)
\begin{eqe}\label{eq:edoLimPlas}
\frac{dy}{dx}=\frac{V_0-v(x,y)}{U_0-u(x,y)},
\end{eqe}%
where $U_0$, $V_0$ are components of the feeding velocity of the die (or extraction velocity at the
output of the die) respectively along the $x$-axis and $y$-axis. One should note that the
conditions (\ref{eq:edoLimPlas}) are reduced to those required on the limits of the plasticity
region in the paper of Czyz \cite{Czyz:1} when $V_0=0$ and that the curves defining the limits
coincide with slip lines (characteristics), that it when we require $dy/dx=\tan\theta(x,y)$ or
$dy/dx=-\cot\theta(x,y)$. Thus the condition (\ref{eq:edoLimPlas}) can be viewed as a relaxation of
the boundary conditions given in the work of Czyz \cite{Czyz:1}. The reason we can use these
relaxed conditions is that we choose the walls of the tool to coincide with the flow lines for a
given solution rather than require the flow of material to be parallel to the walls. Using these
relaxed conditions, we can choose (in some limits) the feeding speed and direction for a tool and
this determines the limits of the plasticity region.
\paragraph*{}The paper is organized as follows. In section \ref{sec:2} we give the infinitesimal generators spanning the Lie algebra
of symmetries $\mathcal{L}$ for the system (\ref{eq:1}) and the discrete transformations leaving it
invariant. A brief discussion on the classification of subalgebras of $\mathcal{L}$ in conjugacy
classes follows. Section \ref{sec:3} is concerned with symmetry reduction. It describes how the
symmetry reduction method (SRM) has been applied to the system (\ref{eq:1}) and the method for
finding partially invariant solutions. More precisely, we give several results obtained from
Ansatzes so that each presented solution
 includes many invariant and partially invariant solutions corresponding to appropriate choices
 of the parameters. We conclude this paper with a discussion on the obtained results and
 some incoming results.
\section{Symmetry algebra and classification of its subalgebras}\label{sec:2}
In this section we study the symmetries of the system (\ref{eq:1}). Following the standard
algorithm \cite{Olver:Application_of_Lie}, the Lie symmetry algebra of the system has been
determined. It is spanned by the eight infinitesimal generators
\begin{gaeq}\label{eq:2}
D_1=x\del_x+y\del_y,\qquad D_2=u\del_u+v\del_v,\qquad B=-y\del_u+x\del_v,\\
P_1=\del_x,\qquad P_2=\del_y,\qquad P_3=\del_\s,\qquad P_4=\del_u,\qquad P_5=\del_v,
\end{gaeq}%
where we use the notation $\del_x=\del/\del_x$, \textit{etc}. The generators $D_1$ and $D_2$
generate dilations respectively in the space of independent variables $\ac{x,y}$ and the space of
dependent variables $\ac{u,v}$. Moreover, $B$ is associated with a kind of boost and the $P_i$,
$i=1,\ldots, 5$ generate translations. The commutation relations for the generators (\ref{eq:2})
are shown in table \ref{tab:1}.
\begin{table}[h]
\caption{Commutation relations for the algebra $\mathcal{L}$.}\label{tab:1}
\begin{center}
\begin{tabular}{|c||c|c|c|c|c|c|c|c|}
  \hline
  $\mathcal{L}$ & $D_1$ & $D_2$ & $B$ & $P_1$& $P_2$ & $P_3$ & $P_4$ & $P_5$ \\
  \hline \hline
  $D_1$ & 0 & 0 & $B$ & $-P_1$ & $-P_2$ & 0 & 0 & 0 \\  \hline
  $D_2$ & 0 & 0 & $-B$ & 0 & 0 & 0 & $-P_4$ & $-P_5$ \\  \hline
  $B$ & $-B$ & $B$ & 0 & $-P_5$ & $P_4$ & 0 & 0 & 0 \\  \hline
  $P_1$ & $P_1$ & 0 & $P_5$ & 0 & 0 & 0 & 0 & 0 \\  \hline
  $P_2$ & $P_2$ & 0 & $-P_4$ & 0 & 0 & 0 & 0 & 0 \\  \hline
  $P_3$ & 0 & 0 & 0 & 0 & 0 & 0 & 0 & 0 \\ \hline
  $P_4$ & 0 & $P_4$ & 0 & 0 & 0 & 0 & 0 & 0 \\  \hline
  $P_5$ & 0 & $P_5$ & 0 & 0 & 0 & 0 & 0 & 0 \\
  \hline
\end{tabular}
\end{center}
\end{table}%
One should note that the system (\ref{eq:1}) is invariant under the discrete transformations:
\begin{aleq}\label{eq:3}
&R_1:\ x\mapsto -x,\quad &&y\mapsto -y,\quad &&\s\mapsto \s, \quad
&&\q\mapsto \q,\quad  &&u\mapsto u,\quad &&v\mapsto v;\\
&R_2:\ x\mapsto x,\quad &&y\mapsto y,\quad &&\s\mapsto \s, \quad &&\q\mapsto \q,\quad  &&u\mapsto
-u,\quad &&v\mapsto -v.
\end{aleq}%
These reflections induce the automorphisms of the Lie algebra $\mathcal{L}$:
\begin{aleq}\label{eq:4}
\mathcal{R}_1:\ &D_1\mapsto D_1,\quad D_2 \mapsto D_2,\quad B\mapsto-B,\quad
P_1\mapsto-P_1,\\
&P_2\mapsto-P_2,\quad P_3\mapsto P_3,\quad P_4\mapsto P_4,\quad P_5\mapsto P_5;\\
\mathcal{R}_2:\ &D_1\mapsto D_1,\quad D_2 \mapsto D_2,\quad B\mapsto-B,\quad
P_1\mapsto P_1,\\
&P_2\mapsto P_2,\quad P_3\mapsto P_3,\quad P_4\mapsto -P_4,\quad P_5\mapsto -P_5.
\end{aleq}%
Since we seek solutions that are invariant and partially invariant of structure defect $\delta=1$,
we only have to classify the subalgebras of codimension 1 and 2. We have used the following
factorization of the Lie algebra $\mathcal{L}$:
\begin{eqe}\label{eq:5}
\mathcal{L}=\ac{\ac{\ac{D_1,D_2}\rhd\ac{B}}\rhd\ac{P_1,P_2,P_4,P_5}}\oplus\ac{P_3},
\end{eqe}%
where $\rhd$ denotes the semi-direct sum and $\oplus$ the direct sum of Lie algebras. The Lie
algebra $\mathcal{L}$ is the direct sum of the center $\ac{P_3}$ with the subalgebra
$\{\{\{D_1,D_2\}\rhd\{B\}\}\rhd\ac{P_1,P_2,P_4,P_5}\}$ which contains an abelian ideal
$\ac{P_1,P_2,P_4,P_5}$. Applying the method \cite{PateraWinter:1,PateraWinter:2,WinterPIEEP}, we
proceed to classify all subalgebras of $\mathcal{L}$ in conjugacy classes under the action of the
automorphisms generated by $G$ and the discrete transformations (\ref{eq:3}). In practice, we can
classify the subalgebras under the automorphisms generated by $G$ and decrease the range of the
parameters that appear in the representative subalgebra of a class using the Lie algebra
automorphisms (\ref{eq:4}). The classification results are shown in table \ref{tab:sadim1} for
subalgebras of codimension 1 and in tables \ref{tab:sadim2a} and \ref{tab:sadim2b} for subalgebras
of codimension 2. For each subalgebra in these tables, a complete set of invariants is given. Table
\ref{tab:sadim2a} contains the codimension 2 subalgebras which admit a symmetry variable, while
table \ref{tab:sadim2b} lists the codimension 2 subalgebras that have no symmetry variable. Despite
the absence of a symmetry variable, the subalgebras in table \ref{tab:sadim2b} lead to PIS.
%
\section{Symmetry reductions and solutions of the reduced systems.}\label{sec:3}%
In this section we use the symmetry reduction method, as presented in
\cite{Olver:Application_of_Lie}, to compute invariant solutions under the action of subgroups of
the symmetry group $G$ of the initial system (\ref{eq:1}). Following the usual reduction procedure
\cite{PW:1}, we consider a subgroup $G_{d,j}\subset G$ associated with a subalgebra
$\mathcal{L}_{d,j}\subset \mathcal{L}$ of dimension $d$. Then the subgroup $G_{d,j}$ admits
$k=p+q-d$ functionally independent invariants $I=(I_1(x,u),\ldots,I_k(x,u))$, where $x$ denotes the
for independent variables and $u$ the dependent variables. We have to consider three different
possibilities.
\begin{itemize}
\item[(i)  ] We first make the following hypotheses:
\begin{itemize}
\item[$\bullet$] $q<d<p+q$,%
\item[$\bullet$] the rank of the Jacobian matrix $(\del I/\del u)$ is $q$,\ \ie $\rank(\del I/\del
u)=q$,%
\item[$\bullet$] the complete set of invariants under the action of $G_{d,j}$ takes the
form\newline $\{\xi_1(x),\ldots,\xi_{k-q}(x), I_1(x,u),\ldots, I_q(x,u)\}$, where $\xi_i$ denotes
the symmetry variables.
\end{itemize}%
With these conditions, the invariant solutions under $G_{d,j}$ are in the form
\begin{eqe}\label{eq:solInv}
u_i=U_i(x,F(\xi)),\qquad i=1,\ldots, q,\qquad F(\xi)=(F_1(\xi),\ldots, F_q(\xi)),
\end{eqe}%
where $U=(U_1,\ldots,U_q)$ is the solution $u=(u_1,\ldots,u_q)$ of the relations
\begin{eqe}\label{eq:relInv}
F_i(\xi)=I_i(x,u),\qquad \xi=(\xi_1(x),\ldots, \xi_{k-q}).
\end{eqe}%
The $F_i$ functions that depend on the $k-q$ variables $\xi_i$, satisfy the differential equation
system that results from the introduction of (\ref{eq:solInv}) into the system (\ref{eq:1}). To
calculate the invariant solutions of the system (\ref{eq:1}), we must consider subalgebras of
codimension $d=1$. Indeed, the system (\ref{eq:1}) is expressed in term of $p=2$ independent
variables, so we cannot reduce it by more than one variable. In other words, we must have $k=5$ to
apply the SRM to
the system (\ref{eq:1}). This is possible only if $d=1$.%
\item[(ii) ] Consider now the situation where $\rank(\del I/\del u)=q'<q$ and $\{\xi_1(x),\ldots$,
$\xi_{k-q'}(x)$, $I_1(x,u),\ldots,I_{q'}(x,u)\}$ is a complete set of invariants of $G_{d,j}$. In
this case, it is only possible to solve the relations
\begin{eqe}\label{eq:relationPI}
F_j(\xi)=I_j(x,u),\quad j=1,\ldots,q',\qquad \xi=(\xi_1,\ldots,\xi_{k-q'}).
\end{eqe}%
for $q'$ of the dependent variables $u$. Without loss of generality, we suppose that we solve the
relations (\ref{eq:relationPI}) for the first $q'$ dependent variables $(u_1,\ldots,u_{q'})$. Then,
introducing
\begin{eqe}\label{eq:solInvPI}
u_i=U_i(x,F(\xi)),\qquad i=1,\ldots, q',\qquad F(\xi)=(F_1(\xi),\ldots, F_{q'}(\xi)),
\end{eqe}%
into the system (\ref{eq:1}), we find a system of differential equations for the functions
$F_i(\xi)$ in which we have the remaining $u_{q'+1}(x),\ldots, u_{q}(x)$, which depend on the
original independent variables $x$. We must add the compatibility conditions on the mixed
derivatives of $u_{q'+1},\ldots, u_{q}$. The obtained solutions are partially invariant. In this
paper, we are interested in PIS of structure defect $\delta=1$ which are obtained from codimension
2 subalgebras. Subalgebras which satisfy the condition $\rank(\del I/\del u)=3<q$  are listed in
table
\ref{tab:sadim2a}.%
\item[(iii)] In the case where the subgroup $G_{d,j}$, corresponding to a codimension 2 subalgebra
$L_{d,j}$, has a complete set of invariants satisfying the condition $\rank \pa{\del I/\del u}=4$,
there always exists, for the considered system (\ref{eq:1}), an invariant in terms of $x,y$ and
$\s$ only. We denote this invariant $\tau$. Consequently, the set of invariants takes the form
$\ac{\tau(x,y,\sigma), I_1(\theta,u,v),I_2(\theta,u,v),I_3(\theta,u,v)}$ and we can solve for
$\theta, u,v$ the relations
\begin{eqe}\label{eq:relInvTau}
F_j(\tau)=I_j(x,u),\qquad i=1,\ldots,3,
\end{eqe}%
and introduce the result in the system (\ref{eq:1}). To justify this particular choice for the
invariant $\tau$ associated to $\mathcal{L}_{d,j}$, one should note that the equations
(\ref{eq:1}.a), (\ref{eq:1}.b), do not involve the quantities $u$,$v$ and that the quantity
$\theta$ is an invariant of the whole Lie algebra $\mathcal{L}$. This choice leaves the equations
(\ref{eq:1}.a), (\ref{eq:1}.b), uncoupled to the equations (\ref{eq:1}.c), (\ref{eq:1}.d).
\end{itemize}%
\paragraph*{}In the case of (i) and (ii), we can always choose the
invariant $I_1=\theta$. Therefore, from the relations (\ref{eq:relationPI}) we have
\begin{eqe}\label{eq:ansatzTheta}
\theta(x,y)=J(\xi(x,y)),
\end{eqe}%
where we write $F_1$ as $J$ for simplification. We will keep this notation for the subsequent
sections of the paper. Introducing $\theta$ given by (\ref{eq:ansatzTheta}) in the equations
(\ref{eq:1}.a), (\ref{eq:1}.b), and using the compatibility conditions on the mixed derivatives of
$\sigma$ relative to $x$, $y$, we obtain an ODE for the function $J$. If we find the solution for
$J$, the compatibility condition is satisfied and then we can integrate by quadrature to find the
solution for $\sigma$. This solution is not necessarily an invariant one, but it includes invariant
solutions for appropriate choices of parameters that appear in $\sigma$.
\paragraph*{}Concerning the case (iii), since there is no symmetry
variable, we propose the solution for $\theta$ in the form
\begin{eqe}\label{eq:ansatzThetaTau}
\theta(x,y)=J(\tau(x,y,\s(x,y))).
\end{eqe}%
This allows one to solve the equations (\ref{eq:1}.a), (\ref{eq:1}.b), for $\theta$, $\sigma$,
without considering the equations (\ref{eq:1}.c) and (\ref{eq:1}.d), because this leaves the
equations uncoupled. None of the codimension 2 subalgebras of table \ref{tab:sadim2b} admit a
symmetry variable, but they possess an invariant in the form $\tau=h_1(x,y) \s + h_2(x,y)$. Thus,
we can express $\sigma $ as
\begin{eqe}\label{eq:formSig}
\s=\frac{\tau-h_2(x,y)}{h_1(x,y)}
\end{eqe}%
and eliminate the quantities $\sigma$ from the equations (\ref{eq:1}.a), (\ref{eq:1}.b), and then
use the compatibility condition on mixed derivatives relative to $x$ and $y$ of the function $\tau$
to obtain an ODE that must be verified by $J$. If we find the solution for $J$, we can find $\tau$
by quadrature and consequently we find $\sigma$ through the relation (\ref{eq:formSig}).
\paragraph*{}Below we present, for each distinct form of the
symmetry variable $\xi$ or of the invariant $\tau$, the most general solution for $\q$ and $\s$ of
the equations (\ref{eq:1}.a), (\ref{eq:1}.b), where we suppose a solution for $\theta$ given by
(\ref{eq:ansatzTheta}) or (\ref{eq:ansatzThetaTau}) depending on the considered case. Then for each
of these solutions, we consider Ansatzes on the form of the velocities $u$ and $v$ to compute
solutions that cover all invariants solutions corresponding to a given form of the symmetry
variable $\xi$ or of the invariant $\tau$. In the table \ref{tab:ansatzs}, the subalgebras
$\mathcal{L}_{i,j}$ are classified by symmetry variable and according to the form of the suggested
Ansatz on the form of the solutions for $u$ and $v$. In general, the solutions are not necessarily
invariant under the action of the symmetry group $G$ but they are so for an appropriate choice of
integration constants. We illustrate with some examples the suggested method.
\subsection{Solution for $\theta$ in the form of a propagation
wave.}\label{sec:3:1} The goal of this section is to construct a solution for the angle $\theta$ in
the form of a propagation wave and to get the corresponding solution for the pressure $\sigma$ and
then, in the subsections \ref{sec:prop_sep_add}, \ref{sec:prop_sep_mul}, we get the velocities $u$
and $v$ from different assumptions made on the form of their solutions,\ \ie we make additive
separation and a multiplicative separation.
\paragraph*{}We look for a solution of the system (\ref{eq:1}.a),
(\ref{eq:1}.b), such that $\q(x,y)$ has the form
\begin{eqe}\label{eq:formSolProp}
\theta(x,y)=J(\xi(x,y)),
\end{eqe}%
with the function $J$ to determine and where
\begin{eqe}\label{eq:8}
\xi(x,y)=a_1 x+ a_2 y,\qquad a_1,a_2\in\mathds{R}.
\end{eqe}%
This type of solution includes, for appropriate values of the parameters $a_1$, $a_2$, all
invariant and partially invariant solutions (for the quantities $\theta$, $\sigma$) corresponding
to subalgebras $L_{1,j}$, $j=1,\ldots,4$ and $L_{2,k}$, $k=1,4,5,6,7,8,10,11$, (that is the
subalgebras listed at line no. 1 in table \ref{tab:ansatzs}). Introducing (\ref{eq:formSolProp}),
(\ref{eq:8}) in (\ref{eq:1}.a), (\ref{eq:1}.b), we get that the following system must be verified
\begin{aleq}\label{eq:9}
&(a)\qquad \s_x(x,y)=2 k \br{a_1 \cos(2J(\xi(x,y)))+a_2\sin(2J(\xi(x,y)))}J'(\xi(x,y)),\\
&(b)\qquad \s_y(x,y)=2 k \br{a_1 \sin(2J(\xi(x,y)))-a_2\cos(2J(\xi(x,y)))}J'(\xi(x,y)).
\end{aleq}%
The compatibility condition on the mixed derivatives for $\sigma$ relative to $x$ and $y$ provides
the next ODE for $J$ in term of $\xi$
\begin{eqe}\label{eq:10}
\br{-\la\sin(2J(\xi))+\mu\cos(2J(\xi))}J''(\xi)+\br{-2\mu\sin(2J(\xi))-2\la\cos(2J(\xi))}J'(\xi)^2=0,
\end{eqe}%
where $c_1,c_2\in\RR$ are integration constants. The solution of (\ref{eq:10}) is
\begin{eqe}\label{eq:11}
J(\xi)=\frac{1-s}{4}\pi+1/2\arctan\pa{\frac{\mu(c_1\xi+c_2)+\la\sqrt{\la^2+\mu^2-(c_1
\xi+c_2)^2}}{\la (c_1\xi + c_2)-\mu \sqrt{\la^2+\mu^2-(c_1 \xi+ c_2)^2}}},
\end{eqe}%
where $s$ denote the sign of the expression $\la (c_1 \xi +c_2)-\mu
\sqrt{\la^2+\mu^2-(c_1\xi+c_2)^2}$ and $\la=a_2^2-a_1^2$, $\mu=-2a_1a_2$. The solution for
$\q(x,y)$ is provided by (\ref{eq:formSolProp}) with $J$ defined by (\ref{eq:11}) and $\xi(x,y)$ by
(\ref{eq:8}). Thereafter, we substitute the solution (\ref{eq:11}) for $J$ in the system
(\ref{eq:9}) and we solve for $\sigma$ by quadrature. We find the solution
\begin{eqe}\label{eq:12}
\s(x,y)=-\frac{c_1 k (a_2 x-a_1 y)}{a_1^2+a_2^2}-\frac{\sqrt{(a_1^2+a_2^2)-\pa{c_1(a_1 x + a_2
y)+c_2}^2}}{a_1^2+a_2^2}+c_3,\quad c_3\in \RR.
\end{eqe}%
We are interested in real solutions, so $\theta$ and $\sigma$ are defined over the domain
\begin{eqe}\label{eq:13}
\W=\ac{(x,y):(a_1^2+a_2^2)-\pa{c_1(a_1 x + a_2 y)+c_2}^2>0}.
\end{eqe}%
Thus, we have that the quantities $\theta$ and $\sigma$ of all invariant solutions for the above
subalgebras are covered by $\theta(x,y)=J(\xi(x,y))$ with $J$ defined by (\ref{eq:11}), $\xi(x,y)$
by (\ref{eq:8}) and $\sigma(x,y)$ by (\ref{eq:12}).
\subsubsection{Additive separation for the velocities $u$ and $v$.}\label{sec:prop_sep_add}
\begin{figure}[h]
\begin{center}
\includegraphics[width=8.5cm]{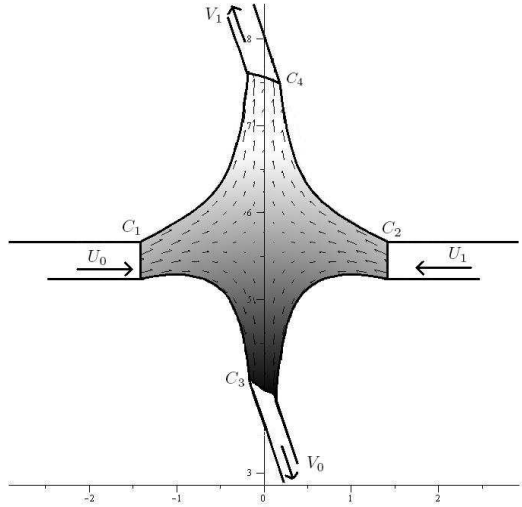}
\end{center}
\caption{Extrusion die corresponding to the solution (\ref{eq:12}), (\ref{eq:31}).} \label{fig:1}
\end{figure}

Corresponding to the solution for $\theta$ and $\sigma$ given respectively by
(\ref{eq:formSolProp}), (\ref{eq:8}), (\ref{eq:11}) and by (\ref{eq:12}), we seek for a solution
$u$ and $v$ of the additive separated form
\begin{eqe}\label{eq:14}
u(x,y)=f(x,y)+F(\xi(x,y)),\qquad v(x,y)=g(x,y)+G(\xi(x,y)).
\end{eqe}%
The first step is to classify the admissible forms for the functions $f$ and $g$. To do this, we
substitute (\ref{eq:14}) in the system (\ref{eq:1}.c), (\ref{eq:1}.d), and we obtain the new system
\begin{aleq}\label{eq:15}
&(a)\qquad
\br{f_y+a_2F'(\xi(x,y))+g_x+a_1G'(\xi(x,y))}\sin(2J(\xi(x,y)))\\
&\qquad\qquad\quad+\br{f_x+a_1F'(\xi(x,y))-g_y-a_2G'(\xi(x,y))}\cos(2J(\xi(x,y)))=0,\\
&(b)\qquad f_x+a_1F'(\xi(x,y))+g_y + a_2 G'(\xi(x,y))=0,
\end{aleq}%
where $F'$, $G'$ represent the derivative of $F$, $G$ relative to $\xi$ respectively. We will keep
this notation in all of the following sections. We next introduce the linear differential operator
\begin{eqe}\label{eq:16}
L_\xi=-a_2P_1+a_1P_2=-a_2\del_x+a_1\del_y,
\end{eqe}%
which annihilates any function in term of the quantity $\xi(x,y)=a_1x+a_2y$. Now, applying it to
the system (\ref{eq:15}) we find the following conditions that constrain $f$ and $g$:
\begin{aleq}\label{eq:17}
&(a)\qquad \pa{L_\xi(f_y+g_x)}\sin(2J(\xi(x,y)))+2(L_\xi
f_x)\cos(2J(\xi(x,y)))=0,\\
&(b)\qquad L_\xi(f_x+g_y)=0.
\end{aleq}%
Then, taking into account that $L_\xi$, $\del_x$ and $\del_y$ commute with each other and using the
notation
\begin{eqe}\label{eq:notafgtilde}
\frak{f}=L_\xi f,\qquad \frak{g}=L_\xi g,
\end{eqe}%
we can rewrite the conditions (\ref{eq:17}) in the new form
\begin{aleq}\label{eq:18}
&(a)\qquad &&\pa{\frak{f}_y+\frak{g}_x}\sin(2J)+2 \frak{f}_x\cos(2J)=0,\\
&(b)\qquad &&\frak{f}_x+\frak{g}_y=0.
\end{aleq}%
We have to consider two cases separately, that is when $\frak{f}$ and $\frak{g}$ verify the
equation (\ref{eq:18}.a) by requiring that the coefficients of the trigonometric functions vanish
\begin{eqe}\label{eq:19}
\qquad\frak{f}_x=\frak{f}_y+\frak{g}_x=0,
\end{eqe}%
or by imposing that  $\frak{f}$ and $\frak{g}$ obey the relation
\begin{eqe}\label{eq:20}
\qquad\frak{f}_y+\frak{g}_x=-2\cot(2J(\xi))\frak{f}_x.
\end{eqe}%
\paragraph*{(i)}If we suppose that (\ref{eq:18}.a) is satisfied by
the conditions (\ref{eq:19}), then we have the following relations
\begin{aleq}\label{eq:23}
&(a)\qquad L_\xi f_x=0,\\
&(b)\qquad L_\xi(f_y+g_x)=0.
\end{aleq}%
If $a_1\neq 0$, $a_2\neq 0$, the solution for $f$ and $g$ of (\ref{eq:23}) takes the form
\begin{eqe}\label{eq:24}
f(x,y)=\zeta_1(\xi)+\w_1y,\qquad g(x,y)=\zeta_2(\xi)+\w_2x,\qquad \w_i\in \RR,\  i=1,2,3.
\end{eqe}%
where the functions $\zeta_i$ are arbitrary functions of $\xi$. If $a_1=0$, then we can set $a_2=1$
without loss of generality. In this case, the functions $f$ and $g$ take the form
\begin{eqe}\label{eq:25}
f(x,y)=\w_1 x+\zeta_1(y),\qquad g(x,y)=\w_2 x + \zeta_2(y),\qquad \w_i\in \RR,\ i=1,2.
\end{eqe}%
where the function $\zeta_i$ are arbitrary functions of $y$. Similarly, we can consider the case
$a_1=1$, $a_2=0$, which leads to the solutions
\begin{eqe}\label{eq:26}
f(x,y)=\zeta_1(x)+\w_1 y, \qquad g(x,y)=\zeta_2(x)+\w_2 y,\qquad \w_i\in \RR,\ i=1,2,
\end{eqe}%
where the functions $\zeta_i$  are arbitrary functions of $x$. Under the hypothesis (\ref{eq:19}),
and according to appropriate values of $a_i$, the form for the functions $f$ and $g$ given by
(\ref{eq:24}), (\ref{eq:25}) or (\ref{eq:26}), are the most general such that the Ansatz
(\ref{eq:14}) reduces the system (\ref{eq:1}.c), (\ref{eq:1}.d), for $u$, $v$ to a system of two
ODE for $F$ and $G$ in term of $\xi$ given by
\begin{aleq}\label{eq:27}
&F'(\xi)=\frac{{a_1(f_x-g_y) +
a_2(f_y+g_x)}\sin(2J(\xi))+2a_2f_x\cos(2J(\xi))}{(a_2^2-a_1^2)\sin(2J(\xi))+2a_1a_2\cos(2J(\xi))},\\
&G'(\xi)=-a_2^{-1}\pa{a_1F'(\xi)+f_x+g_y}.
\end{aleq}%
We distinguish two different cases for the solutions $F$ and $G$ of (\ref{eq:27}), that is when
$a_1\neq0$ and $a_2\neq0$ or when $a_2=0$.
\paragraph*{(a)}If $a_1\neq 0$ and $a_2\neq 0$, we obtain
\begin{eqe}\label{eq:29}
F(\xi)=-\zeta_1(\xi)+\frac{a_1(\omega_1+\omega_2)\cos(2J(\xi))}{2c_1}+\frac{a_2\omega_1\sin(2J(\xi))}{c_1}+c_4,
\end{eqe}%
and
\begin{eqe}\label{eq:30}
G(\xi)=-\zeta_2(\xi)-\frac{a_1^2(\omega_1+\omega_2)\cos(2J(\xi))}{c_1}-\frac{a_1\omega_1\sin(2J(\xi))}{c_1}-\frac{\omega_1+\omega_2}{a_2}\xi+c_5.
\end{eqe}%
Redefining the parameters $\w_1$, $\omega_2$, the solutions $u$ and $v$ of (\ref{eq:1}.c),
(\ref{eq:1}.d), are then
\begin{aleq}\label{eq:31}
&(a)\qquad u(x,y)=c_4 x+\frac{a_1(c_4+c_5)\cos(2J(\xi(x,y)))}{2c_1}+\frac{a_2c_4\sin(2(J(\xi(x,y))))}{c_1}+c_6,\\
&(b)\qquad
v(x,y)=-\frac{a_1^2(c_4+c_5)\cos(2J(\xi))}{2a_2c_1}-\frac{a_1c_4\sin(2J(\xi))}{c_1}-\frac{a_1(c_4+c_5)x}{a_2}-c_4
y+c_7,
\end{aleq}%
where the $c_i$ are integration constants. So, we obtain a solution of the system (\ref{eq:1}) by
defining the angle $\theta$ by (\ref{eq:formSolProp}), (\ref{eq:11}), the pressure $\sigma$ by
(\ref{eq:12}) and the velocities $u$ and $v$ by (\ref{eq:31}), with $\xi(x,y)=a_1 x+a_2 y$. As can
be seen in the illustration, we plotted in figure \ref{fig:1} the shape of an extrusion die
corresponding to the solution (\ref{eq:31}) for the velocities $u$ and $v$. The vector field
defined by these velocities is traced inside the tool. The mean pressure in the extrusion die,
which is given by (\ref{eq:12}), is represented by shades of grey, from pale grey for lower
pressure to dark grey for higher pressure. The parameters have been set to $k=0.0027$, $a_1=1$,
$a_2=1$, $c_1=0,5$, $c_2=0$, $c_3=0$, $c_4=0$, $c_5=-1$, $c_6=0$, $c_7=0$ and the feeding
velocities modulus $U_0$, $U_1$, are both equal to $\sqrt{2}$, while the extraction velocities
modulus $V_0$, $V_1$ are both equal to $\sqrt{3}$. The curves $C_1$, $C_2$ determine the limit of
the plasticity region at the mouth of the die while the curves $C_3$, $C_4$, have the same meaning
at the end of the die. It's a double feeding tool which has a symmetric configuration under the
reflection $x\mapsto -x$ except near the plasticity limits defined by $C_3$ and $C_4$. Other
feeding and extraction settings are possible, that is the angle and the velocities modulus, but no
setting with completely vertical extraction speeds were found.
\paragraph*{(b)}Now, if $a_1=0$ and $a_2=1$, then $f$ and $g$ are
given by (\ref{eq:25}). In this case, we find
\begin{aleq}\label{eq:32}
&F(y)=-\zeta_1(y)-\w_2 y -2\w_1\int\cot 2J(y)dy+c_4,\\
&G(y)=-\w_1 y - \zeta_2(y)+c_5.
\end{aleq}%
By redefining $\omega_1$, $\omega_2$, the solution may be written as
\begin{aleq}\label{eq:33}
&u(x,y)=c_4 x - c_5 y-2c_4\int \cot2J(y)dy+c_5,\\
&v(x,y)=-c_4 y+c_7,
\end{aleq}%
where the $c_i$ are real arbitrary constants. So, a solution of the system (\ref{eq:1}) consists of
the angle $\theta$ by (\ref{eq:formSolProp}), (\ref{eq:11}), the pressure $\sigma$ by (\ref{eq:12})
and the velocities $u$ and $v$ by (\ref{eq:33}), with $\xi(x,y)=y$. For this solution, we did not
trace an extrusion die because of the similarity with the next case where a tool will be plotted.
\paragraph*{}If $a_1=1$ and $a_2=0$, then the functions $f$ and $g$
are given by (\ref{eq:26}). We find in this case
\begin{aleq}\label{eq:34}
&F(x)=-\zeta_1(x)-\w_2 x+c_4,\\
&G(x)=-\w_1 x-\zeta_2(x)+2\w_2\int \cot2J(x)dx+c_5.
\end{aleq}%
Redefining $\omega_1$, $\omega_2$, the solution for $u$ and $v$ of (\ref{eq:1}.c), (\ref{eq:1}.d)
is
\begin{aleq}\label{eq:35}
&u(x,y)=-c_5 x+c_4 y+c_6,\\
&v(x,y)=-c_4 x+c_5 y +2c_5 \int\cot2J(x)dx+c_7,
\end{aleq}%
where the $c_i$ are real arbitrary constants. The system (\ref{eq:1}) is solved by the angle
$\theta$ by (\ref{eq:formSolProp}), (\ref{eq:11}), the mean pressure $\sigma$ by (\ref{eq:12}) and
the velocities $u$ and $v$ by  (\ref{eq:35}), with $\xi(x,y)=x$.
\begin{figure}[h]
\begin{center}
\includegraphics[width=8.5cm]{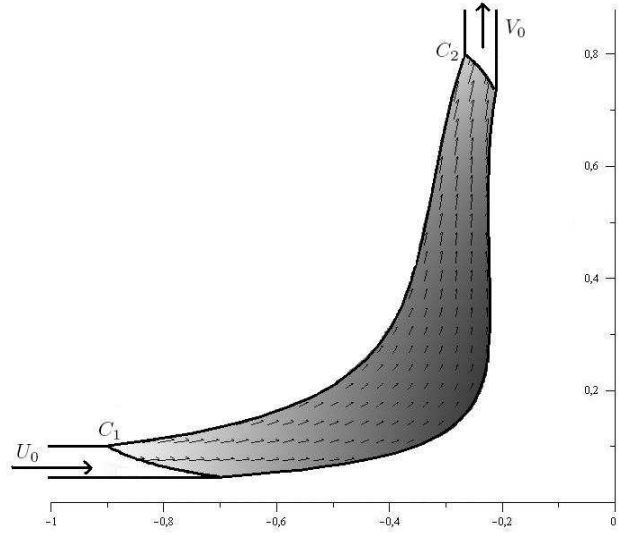}
\end{center}
\caption{Extrusion die corresponding to the solution (\ref{eq:12}), (\ref{eq:31}).}%
\label{fig:2}
\end{figure}
An extrusion die corresponding to the solution (\ref{eq:35}), with the mean pressure provided by
(\ref{eq:12}), is shown in figure \ref{fig:2} for parameter values set to $k=0.0027$, $a_1=1$,
$a_2=0$, $c_1=0,5$, $c_2=0$, $c_3=0$, $c_4=-1$, $c_5=1$, $c_6=2$, $c_7=0$. The pressure is
represented by the shades of grey ranging from pale to dark as the pressure increase. The curves
$C_1$ and $C_2$ define the plasticity regions respectively at the mouth and at the end of the tool.
The feeding velocity used is $U_0=1$ and the extraction velocity is $V_0=1$. For example, this die
would be used to curve a plate or a rectangular rod.
\paragraph*{(ii)}Now, we consider the case where $\frak{f}$ and
$\frak{g}$ satisfy the condition (\ref{eq:20}). By the use of (\ref{eq:18}), the compatibility
condition on $\frak{g}$ leads to the hyperbolic equation
\begin{eqe}\label{eq:36}
\frak{f}_{xx}-\frak{f}_{yy}-2\cot(2J(\xi)) \frak{f}_{xy}+4 a_2 \csc(2J(\xi)) J'(\xi) \frak{f}_x=0,
\end{eqe}%
over all the domain $\W$ defined by (\ref{eq:13}). So, we introduce the variables change  $(x,y)$
to $(\vf,\psi)$ defined by
\begin{aleq}\label{eq:37}
&\vf=x+\frac{1}{c_1}\pa{a_1\cos(2J)+a_2\sin(2J)-2J},\\
&\psi=x+\frac{1}{c_1}\pa{a_1\cos(2J)+a_2\sin(2J)+2J},
\end{aleq}%
which enables us to rewrite (\ref{eq:36}) in the simplified form
\begin{eqe}\label{eq:37b}
\frak{f}_{\vf\psi}=\frac{c_1}{4a_2}(\frak{f}_\vf-\frak{f}_\psi)\cot(2J)+\frac{c_1}{4a_2}\frac{\frak{f}_\vf+\frak{f}_\psi}{\sin(2J)}.
\end{eqe}%
One should note that taking the difference of equations (\ref{eq:37}), we deduce
\begin{eqe}\label{eq:37c}
J=\frac{c_1(\psi-\vf)}{4a_2}.
\end{eqe}%
We can find a particular solution of (\ref{eq:37b}) by assuming the solution in the separated form
$\frak{f}(\vf,\psi)=P(K)+Q(J)$ where we use $K=\frac{c_1(\vf+\psi)}{4a_2}$. Under such a
hypothesis, the solution of the equation (\ref{eq:37}) is
\begin{eqe}\label{eq:38}
\frak{f}(\vf,\psi)=\frac{\w_1
c_1}{4a_2}(\vf+\psi)+\w_2\cos\pa{\frac{c_1(\psi-\vf)}{2a_2}}+\frac{\w_1}{2}\sin\pa{\frac{c_1(\psi-\vf)}{2a_2}}+\w_3,
\end{eqe}%
where the $\w_i\in \RR,\ i=1,\ldots,3$, are integration constants. By considering the change of
variable (\ref{eq:37}), the system (\ref{eq:18}) takes the form
\begin{eqe}\label{eq:39}
\frak{g}_\vf=-\frac{\cos\pa{\frac{c_1(\psi-\vf)}{4a_2}}}{\sin\pa{\frac{c_1(\psi-\vf)}{4a_2}}}
\frak{f}_\vf,\qquad
\frak{g}_\psi=\frac{\sin\pa{\frac{c_1(\psi-\vf)}{4a_2}}}{\cos\pa{\frac{c_1(\psi-\vf)}{4a_2}}}
\frak{f}_\psi,
\end{eqe}%
from which we find the solution for  $\frak{g}$ defined by
\begin{eqe}\label{eq:40}
\frak{g}(\vf,\psi)=-\frac{\w_1}{2}\cos\pa{\frac{c_1 (\vf-\psi)}{2a_2}}+\w_2\sin\pa{\frac{c_1
(\vf-\psi)}{2a_2}}-\frac{c_1\w_2}{2a_2}(\vf+\psi)+\w_4.
\end{eqe}%
where $\w_i\in \RR,\ i=1,\ldots,4$. Finally, returning to the initial variables and considering the
relations (\ref{eq:notafgtilde}), we obtain $f$ and $g$ defined by
\begin{aleq}\label{eq:41}
f(x,y)=&\frac{\w_1c_1 (2a_1\xi(x,y)-a_2\eta(x,y))\eta(x,y)}{4 a_2
(a_1^2+a_2^2)}+\left(\pa{\w_2+\frac{\w_1a_1}{2a_2}}\cos(2J(\xi(x,y)))\right.\\
&\qquad\left.+\w_1\sin(2J(\xi(x,y)))\right)\eta(x,y)+\w_3\eta(x,y)+\zeta_1(\xi(x,y)),\\
g(x,y)=&\frac{-\w_2 c_1 (2a_1\xi(x,y)-a_2\eta(x,y))\eta(x,y)}{2 a_2
(a_1^2+a_2^2)}-\left(\pa{\frac{\w_1}{2}+\frac{\w_2
a_1}{2}}\cos(2J(\xi(x,y)))\right)\eta(x,y)\\
&\qquad+\w_4\eta(x,y)+\zeta_2(\xi(x,y)),
\end{aleq}%
where we introduced the notation $$\xi(x,y)=a_1x+a_2y,\qquad\eta(x,y)=-a_2 x+a_1y$$%
and where $\zeta_i(\xi)$ are real arbitrary functions of one variable.
\paragraph*{}It remains for us to integrate the equations (\ref{eq:15})
with $f$ and $g$ given by (\ref{eq:41}), which form a system of two ODE for the unknown quantities
$F$ and $G$ in term of $\xi$. The solution for $F$ is expressed by the quadrature
\begin{aleq}\label{eq:42}
F(\xi)=&\int\br{2 a_2\kappa_3 \pa{(a_2-1)\sin(2J)+2a_1\cos(2J)}}^{-1}\\
&\times\Bigg[\w_1\Big(-a_1c_1\pa{\kappa_1\sin(2J(\xi))-2a_2\cos(2J(\xi))}\xi
-\kappa_3\Big(2a_2\kappa_1\sin^2(2J(\xi))\\
&+(a_1+3a_2)\kappa_2\cos(2J(\xi))\sin(2J(\xi))-2a_1a_2\cos^2(2J(\xi))\Big)\Big)\\
&+\w_2\Big(-a_1c_1\kappa_1\xi\sin(2J(\xi))-2\kappa_3\cos(2J(\xi))\pa{\kappa_1^2\sin(2J(\xi))-2a_2^2\cos(2J(\xi))}\Big)\\
&\quad
-2\w_3a_2\kappa_3\pa{\kappa_1\sin(2J(\xi))-2a_2\cos(2J(\xi))}+2\w_4a_2\kappa_1\kappa_3\sin(2J(\xi))\\
&+2\kappa_3a_2\Big(\pa{\kappa_2\sin(2J(\xi))-2a_1\cos(2J(\xi))}\frac{d\zeta_1}{d\xi}-\kappa_2\sin(2J(\xi))\frac{d\zeta_2}{d\xi}\Big)\Bigg]d\xi+c_4.
\end{aleq}%
It's the same for the solution for $G$,
\begin{aleq}\label{eq:43}
G(\xi)=&\int\br{2a_1a_2\kappa_3\pa{(a_2-1)\sin(2J(\xi))+2a_1\cos(2J(\xi))}}^{-1}\\
&\times\Bigg[\w_1\Big(a_1c_1\xi\big(\pa{a_1+a_2^2}\sin(2J(\xi))\\
&+2a_2(a_1-1)\cos(2J(\xi))\big)+\kappa_3\big(2a_2(a_1+a_2^3)\sin^2(2J(\xi))\\
&+\pa{(6a_1-3)a_2^2+a_1^2}\big)\sin(2J(\xi))\cos(2J(\xi))+2a_1a_2(2a_1-1)\cos^2(2J(\xi))\Big)\\
&+\w_2\Big(2c_1a_1\pa{a_2(a_1+1)\sin(2J(\xi))+2
a_1^2\cos(2J(\xi))}\xi\\
&+2\kappa_3\cos(2J(\xi))\big(a_2(a_1^2+2a_1+a_2^2)\sin(2J(\xi))\pa{2(a_1-1)a_2^2+2a_1^3}\cos(2J(\xi))\big)\Big)\\
&+2\w_3a_2\kappa_3\Big((a_1+a_2^2)\sin(2J(\xi))+2a_2(a_1-1)\cos(2J(\xi))\Big)\\
&-2\w_4a_2\kappa_3\Big(a_2(a_1+1)\sin(2J(\xi))+2a_1^2\cos(2J(\xi))\Big)\\
&-2\kappa_3a_2(a_1-1)\Big(a_2\sin(2J(\xi))+2a_1\cos(2(2J(\xi)))\Big)\frac{d\zeta_1}{d\xi}\\
&+\Big((a_2^2-a_1)\sin(2J(\xi))+2a_1a_2\cos(2J(\xi))\Big)\frac{d\zeta_2}{d\xi}\Bigg]d\xi +c_5,
\end{aleq}%
where we used $\kappa_1=a_1+a_2$, $\kappa_2=a_1-a_2$, $\kappa_3=a_1^2+a_2^2$. The functions
$\zeta_i$ being arbitrary, they could be chosen in a way such that their derivatives $\del
\zeta_i/\del \xi$ simplify the integration procedure for $F$ and $G$. Finally, the solutions for
$u$ and $v$ are
\begin{eqe}\label{eq:43b}
u(x,y)=f(x,y)+F(a_1 x+a_2 y),\qquad v(x,y)=g(x,y)+G(a_1 x+a_2y),
\end{eqe}%
where the functions $f$ and $g$ are given by (\ref{eq:41}) while $F$ and $G$ are given respectively
by the quadratures (\ref{eq:42}) and (\ref{eq:43}). So, we get a solution of the system
(\ref{eq:1}) by defining the angle $\theta$ by (\ref{eq:formSolProp}), (\ref{eq:11}), the mean
pressure $\sigma$ by (\ref{eq:12}) and the velocities $u$ and $v$ by (\ref{eq:43b}), with
$\xi(x,y)=a_1x+a_2y$.
\subsubsection{Multiplicative separation for the velocities $u$ and $v$.}\label{sec:prop_sep_mul}
\begin{figure}[h]
\begin{center}
\includegraphics[width=8.5cm]{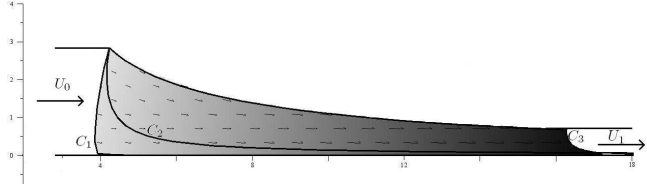}
\end{center}
\caption{Extrusion die corresponding to the solution (\ref{eq:12}), (\ref{eq:70}).}%
\label{fig:3}
\end{figure}

Considering the solution of (\ref{eq:1}.a), (\ref{eq:1}.b) for the angle $\theta$ provided by
(\ref{eq:formSolProp}), (\ref{eq:8}), (\ref{eq:11}), and for the pressure $\sigma$ by
(\ref{eq:12}), we now propose the solution for $u$ and $v$ to the equations (\ref{eq:1}.c),
(\ref{eq:1}.d), in the multiplicative separated form
\begin{eqe}\label{eq:44}
u(x,y)=f(x,y)F(\xi(x,y)),\qquad v(x,y)=g(x,y)G(\xi(x,y)),
\end{eqe}%
where we suppose $f\neq 0$ and $g\neq 0$. For classifying the admissible $f$ and $g$, we substitute
the velocities (\ref{eq:44}) into the system (\ref{eq:1}.c) and (\ref{eq:1}.d). This gives
\begin{aleq}\label{eq:45}
&(a)\qquad \pa{f_yF(\xi(x,y))+a_2f F'(\xi(x,y))+g_xG(\xi(x,y))+a_1 g G'(\xi(x,y))}\sin2J(\xi(x,y))+\\
&\qquad \qquad +\pa{f_xF(\xi(x,y))+a_1fF'(\xi(x,y))-g_yG(\xi)-a_2 g G'(\xi)}\cos2J(\xi),\\
&(b)\qquad f_x F(\xi(x,y))+ a_1 f F'(\xi(x,y))+g_yG(\xi(x,y))+a_2
 gG'(\xi(x,y))=0.
\end{aleq}%
Applying to (\ref{eq:45}.b) the linear differential operator $L_{\xi}$ defined by (\ref{eq:16}), we
have that the following equation must be satisfied
\begin{eqe}\label{eq:46}
L_{\xi}\pa{\frac{f_x}{g}}F+a_1 L_{\xi}\pa{\frac{f}{g}}F'+L_{\xi}\pa{\frac{g_y}{g}}G=0,
\end{eqe}%
where we have $F=F(\xi(x,y))$, $G=G(\xi(x,y))$. In dealing with equation (\ref{eq:46}), there are
three different cases to consider:
\begin{aleq}\label{eq:47}
&(i)\qquad &&
L_{\xi}\pa{\frac{f_x}{g}}=L_{\xi}\pa{\frac{f}{g}}=L_{\xi}\pa{\frac{g_y}{g}}=0,\\
&(ii)\qquad && L_{\xi}\pa{\frac{g_y}{g}}=0,\qquad
L_{\xi}\pa{\frac{f}{g}}\neq 0,\qquad L_{\xi}\pa{\frac{f_x}{g}}\neq 0\\
&(iii) &&L_{\xi}\pa{\frac{g_y}{g}}\neq 0.
\end{aleq}%
\paragraph*{(i)}In the first case, the functions $f$ and $g$ must
take the form
\begin{eqe}\label{eq:48}
f(x,y)=e^{\w_1 x} \zeta_2(\xi),\qquad g(x,y)= e^{\w_1 x}\zeta_1(\xi),\qquad \w_1\in\RR,
\end{eqe}%
where $\omega_1$ corresponds to a damping factor when $\omega_1<0$. Introducing (\ref{eq:48}) in
the system (\ref{eq:45}), we reduce this one to an ODE system for the functions $F$ and $G$ in term
of $\xi$ from which we find
\begin{eqe}\label{eq:49}
G(\xi)=A_1(\xi) F(\xi)+A_2(\xi) F'(\xi),
\end{eqe}%
where the coefficients $A_i(\xi)$, $i=1,2$, are given by
\begin{aleq}\label{eq:49b}
&A_1(\xi)=\frac{1}{a_2\w_1}\frac{\zeta_2(\xi)}{\zeta_1(\xi)}\pa{\pa{-2\w_1a_2 +\lambda \frac{\zeta_2'}{\zeta_2}}\cot(2J(\xi))-\pa{\mu \frac{\zeta_2'}{\zeta_2}-a_1\w_1}},\\
&A_2(\xi)=\frac{1}{a_2\w_1}\frac{\zeta_2(\xi)}{\zeta_1(\xi)}\pa{\lambda \cot(2J(\xi))-\mu},
\end{aleq}%
where $\zeta_1$, $\zeta_2$ are arbitrary functions of one variable, while the function $F$ must
satisfy the ODE
\begin{aleq}\label{eq:50}
&F''(\xi)+\frac{\w_1 \zeta_2(\xi)+a_1\zeta_2'(\xi)+a_2\zeta_1'(\xi)A_1(\xi)+a_2\zeta_1(\xi)
A_1'(\xi)}{a_2
\zeta_1(\xi)A_2(\xi)}F'(\xi)\\
&+\frac{a_1\zeta_2(\xi)+a_2\zeta_1'(\xi)A_2(\xi)+a_2\zeta_1(\xi)(A_1(\xi)+A_2'(\xi))}{a_2
\zeta_1(\xi)A_2(\xi)}F(\xi)=0.
\end{aleq}%
If we find the solution for $F$ of the equation (\ref{eq:50}) for a particular choice of the
functions $\zeta_1$ and $\zeta_2$, then the solutions for the system (\ref{eq:1}.c),
(\ref{eq:1}.d), for $u$ and $v$ will be given by (\ref{eq:44}) with $f$ and $g$ defined by
(\ref{eq:48}) and $G$ by (\ref{eq:49}). However, by imposing $F(\xi)=\exp(K(\xi))$, the problem of
solving (\ref{eq:50}) is reduced to the one of solving the Ricatti equation
\begin{eqe}\label{eq:50m}
K''(\xi)+K'(\xi)^2+B_1(\xi) K'(\xi)+B_2(\xi)=0,
\end{eqe}%
where
\begin{aleq*}
B_1(\xi)=&\frac{8a_1a_2J'(\xi)^2}{c_1\sin(2J(\xi))}+\frac{2\zeta_2'(\xi)}{\zeta_2(\xi)}+\frac{4\omega_1\pa{a_1\sin(2J(\xi))-a_2\cos(2J(\xi))}J'(\xi)}{c_1},\\
B_2(\xi)=&\frac{\zeta''_2(\xi)}{\zeta_2(\xi)}+\frac{4J'(\xi)\pa{2a_1a_2J'(\xi)+\omega_1a_1\sin(2J(\xi))^2-\omega_1a_2\sin(2J(\xi))\cos(2J(\xi))}\zeta_2'(\xi)}{c_1\sin(2J(\xi))\zeta_2(\xi)}\\
&+\frac{2\omega_1J'(\xi)\pa{4a_2J'(\xi)+\omega_1\sin(2J(\xi))^2}}{c_1\sin(2J(\xi))}.
\end{aleq*}%
If we choose the function $\zeta_2$ to be a particular solution of the Ricatti equation obtained by
imposing that the coefficient $B_2(\xi)\equiv 0$, then the ODE (\ref{eq:50m}) becomes a nonlinear
Bernoulli equation for $K'(\xi)$ and the solution for $K(\xi)$ is obtained by quadrature. In this
case, the solution for $F$ takes the form
\begin{eqe}\label{eq:50m2}
F(\xi)=c_4\int \exp\pa{-\int B_1(\xi) d\xi} d\xi+c_5, \qquad c_4,c_5\in\RR.
\end{eqe}%
We substitute (\ref{eq:50m2}) in (\ref{eq:49}) and calculate the  following solution for $G$:
\begin{eqe}\label{eq:50m3}
G(\xi)=c_4\br{A_1(\xi) \int \exp\pa{-\int B_1(\xi)d\xi}d\xi+A_2(\xi)\exp\pa{-\int
B_1(\xi)d\xi}}+c_5A_2(\xi).
\end{eqe}%
Finally, the velocities $u$ and $v$ are
\begin{aleq}\label{eq:50m4}
&u(x,y)=e^{\omega_1 x}\zeta_2(y/x)\pa{c_4\int \exp\pa{-\int^{\xi=y/x} B_1(\xi)d\xi}d\xi+c_5},\\
&v(x,y)=e^{\omega_1x}\zeta_1(y/x)\pa{c_4\pa{A_1(y/x)\int^{\xi=y/x} e^{-\int
B_1(\xi)d\xi}d\xi+A_2(\xi)e^{-\int B_1(\xi)d\xi}}+c_5A_1(\xi)},
\end{aleq}%
where $A_1$, $A_2$ are defined by (\ref{eq:49b}), $\zeta_1(\xi)$ is an arbitrary function and
$\zeta_2(\xi)$ is a solution of the Ricatti equation
\begin{aleq*}
&\zeta''_2(\xi)+\frac{4J'(\xi)\pa{2a_1a_2J'(\xi)+\omega_1a_1\sin(2J(\xi))^2-\omega_1a_2\sin(2J(\xi))\cos(2J(\xi))}\zeta_2'(\xi)}{c_1\sin(2J(\xi))}\\
&+\frac{2\omega_1J'(\xi)\pa{4a_2J'(\xi)+\omega_1\sin(2J(\xi))^2}}{c_1\sin(2J(\xi))} \zeta_2(\xi)=0.
\end{aleq*}%
So, we have obtained a solution of the system (\ref{eq:1}) by redefining the angle $\theta$ by
(\ref{eq:formSolProp}), (\ref{eq:11}), the pressure $\sigma$ by (\ref{eq:12}) and the velocities
$u$ and $v$ by (\ref{eq:50m4}), with $\xi(x,y)=a_1x+a_2y$.
\paragraph*{(ii)}In the second case, we have that
\begin{eqe}\label{eq:51}
g(x,y)=\w_4 e^{\w_3 x}\zeta_1(\xi),\qquad \w_3,\w_4\in \RR,
\end{eqe}%
and the function $f(x,y)$ must be a solution of the system of two PDE
\begin{aleq}\label{eq:52}
f_x=&-a_1\zeta_2'(\xi)f+g\zeta_3(\xi)\\
f_y=&-a_2\zeta_2'(\xi)f-\w_3\w_4 e^{\w_3 x}
\pa{\frac{\w_2}{\w_1}e^{-\zeta_2(\xi)}-\frac{\w_1}{a_2}\int\zeta_1(\xi)\zeta_3(\xi)e^{\zeta_2(\xi)}d\xi}\\
&+\w_4 e^{\w_3 x}\zeta_1(\xi)\zeta_3(\xi) \frac{-2a_2\cos(2J(\xi))+a_1\sin(2J(\xi))}{a_2
\sin(2J(\xi))},
\end{aleq}%
where the $\w_i$, $i=1,2,3,4$, are real parameters and, in order that the system (\ref{eq:52}) be
compatible, the functions $\zeta_i(\xi)$, $i=1,2,3$, must obey the condition
\begin{aleq}\label{eq:53}
&2\zeta_1(\xi)\zeta_3(\xi)\pa{\frac{2a_1
a_2J'(\xi)}{\sin(2J(\xi))}-\w_3\pa{-a_1\sin(2J(\xi))+a_2\cos(2J(\xi))}}\\
&+\pa{\la \sin(2J(\xi))+\mu
\cos(2J(\xi))}\pa{\zeta_1(\xi)\zeta'_3(\xi)+\zeta_1(\xi)\zeta_3(\xi)\zeta'_2(\xi)+\zeta_1(\xi)\zeta'_1(\xi)}\\
&+\frac{a_2\w_3^2}{\w_1}e^{-\zeta_2(\xi)}\sin(2J(\xi))\pa{-\w_2+\frac{\w_1}{a_2}\int\zeta_1(\xi)\zeta_3(\xi)e^{\zeta_2(\xi)}d\xi}=0.
\end{aleq}%
The functions $F$ and $G$ are given by
\begin{eqe}\label{eq:54}
F(\xi)=\w_1e^{\zeta_2(\xi)},\qquad G(\xi)=\frac{1}{\zeta_2(\xi)}\pa{\frac{-\w_1}{a_2}\int
\zeta_1(\xi)\zeta_3(\xi)e^{\zeta_2(\xi)}d\xi}.
\end{eqe}%
\paragraph*{}As an illustration, we compute an explicit solution of the
system (\ref{eq:1}). To do this, we choose
\begin{eqe}\label{eq:55}
\zeta_2(\xi)=-\ln(\zeta_1(\xi)),\qquad \zeta_3(\xi)=2 \w_5 c_1^{-1} \sin(2J(\xi)) J'(\xi),\qquad
\w_5\in\RR,
\end{eqe}%
so that the equation (\ref{eq:53}) is verified and consequently the system (\ref{eq:52}) can be
solved because it becomes compatible for $f$. The solution of the system (\ref{eq:52}) is then
\begin{eqe}\label{eq:56}
f(x,y)=c_1^{-1}\pa{2\w_4\w_5\int J'(\xi)\sin(2J(\xi))dx+\frac{\w_4\w_5}{a_1a_2}y+\w_7}\zeta_1(\xi),
\end{eqe}%
where $\xi=a_1 x+a_2 y$ and the function $\zeta_1(\xi)$ has as its only restriction that its first
derivative $d\zeta_1(\xi)/d\xi\neq 0$.
\paragraph*{}Finally, by introducing (\ref{eq:51}), (\ref{eq:54})
and (\ref{eq:56}) in the expression (\ref{eq:44}), with the function $\theta(x,y)$ defined by
(\ref{eq:formSolProp}), (\ref{eq:8}) and (\ref{eq:11}), the solutions for $u$ and $v$ of the system
(\ref{eq:1}.c), (\ref{eq:1}.d), are
\begin{aleq}\label{eq:57}
&u(x,y)=-c_4a_2\cos(2J(\xi))+c_1c_4y+c_5,\\
&v(x,y)=c_4a_1\cos(2J(\xi))+c_6,\qquad c_i\in\RR,\ i=1,2,3,4,
\end{aleq}%
where the real parameters $\w_i$ have been redefined to simplify the expression (\ref{eq:57}) in
which the new parameters are $c_i$. So, we have a solution of the system (\ref{eq:1}) by defining
the angle $\theta$ by (\ref{eq:formSolProp}), (\ref{eq:11}), the pressure $\sigma$ by (\ref{eq:12})
and the velocities $u$ and $v$ by (\ref{eq:57}), with $\xi(x,y)=a_1x+a_2y$.
\paragraph*{(iii)}If the condition  (\ref{eq:47}.iii) is satisfied,
then we have
\begin{eqe}\label{eq:58}
F(\xi)=\w_1 e^{-\zeta_1(\xi)/a_1},\qquad G(\xi)=-\w_1\zeta_2(\xi)e^{-\zeta_1(\xi)/a_1},\quad
\w_1\in\RR,
\end{eqe}%
when the functions $f(x,y)$ and $g(x,y)$ are solutions of the system
\begin{aleq}\label{eq:59}
f_x=&\zeta'_1(\xi)f+\zeta_2(\xi) g_y +f\zeta_3(\xi)
g,\\
f_y=&\frac{a_2}{a_1}\zeta'_1(\xi)f+\zeta_2(\xi)g_x-2\cot(2J(\xi))
\zeta_2(\xi)g_y\\
&-\pa{a_1a_2^{-1}-2\cot(2J(\xi))}\zeta_3(\xi) g,
\end{aleq}%
where the functions $\zeta_1(\xi)$, $\zeta_2(\xi)$ are arbitrary and
\begin{eqe}\label{eq:60}
\zeta_3(\xi)=a_2\pa{-\frac{\zeta_2(\xi)\zeta'_1(\xi)}{a_1}+\zeta'_2(\xi)}.
\end{eqe}%
Using the compatibility condition on the mixed derivatives of the function $f(x,y)$ relative to $x$
and $y$, we obtain the second order ODE for $g(x,y)$
\begin{aleq}\label{eq:61}
&\zeta_2(\xi)\pa{g_{xx}-g_{yy}-2\cot(2J(\xi))g_{xy}}-2\zeta_3(\xi)\pa{a_1a_2^{-1}-\cot(2J(\xi))}g_x\\
&+2\pa{-\zeta_3(\xi)\pa{1+a_1a_2^{-1}\cot(2J(\xi))}+\frac{2a_1\zeta_2(\xi)J'(\xi)}{\sin^2(2J(\xi))}}g_y\\
&+\br{\zeta_3(\xi)\pa{\frac{4a_1J'(\xi)}{\sin^2(2J(\xi))}+\zeta'_1(\xi)\frac{\la-\mu\cot(2J(\xi))}{a_1a_2}}-\zeta'_3(\xi)\pa{\frac{\la-\mu\cot(2J(\xi))}{a_2}}}g=0.
\end{aleq}%
It's a hyperbolic equation on the domain $\Omega$ defined by (\ref{eq:13}). So the change of
variable
\begin{aleq}\label{eq:62}
&(a)\qquad&&\phi(x,y)=x-\frac{2a_2J(\xi)}{c_1}+\frac{a_1(\cos(2J(\xi))+1)+a_2\sin(2J(\xi))}{c_1},\\
&(b)&&\psi(x,y)=x+\frac{2a_2J(\xi)}{c_1}+\frac{a_1(\cos(2J(\xi))+1)+a_2\sin(2J(\xi))}{c_1},
\end{aleq}%
transforms the equation (\ref{eq:61}) to the simplified form
\begin{aleq}\label{eq:63}
&0=g_{\phi\psi}(\phi,\psi)+\frac{c_1}{4a_2}\pa{-\frac{\cos(2J(\xi))-1}{\sin(2J(\xi))}+\frac{\zeta_3(\xi)}{a_2\zeta_2(\xi)J'(\xi)}}
g_{\phi}(\phi,\psi)\\
&+\frac{c_1}{4a_2}\pa{\frac{\cos(2J(\xi))+1}{\sin(2J(\xi))}-\frac{\zeta_3(\xi)}{a_2\zeta_2(\xi)J'(\xi)}}g_{\psi}(\phi,\psi)\\
&-\frac{c_1}{2a_2^2}\pa{\frac{a_1\zeta_3(\xi)}{\zeta_2(\xi)\sin(2J(\xi))}+\frac{\la\sin(2J(\xi))-\mu\cos(2J(\xi))}{4a_2a_1\zeta_2(\xi)J'(\xi)}\pa{\zeta_3(\xi)\zeta'_1(\xi)-a_1\zeta'_3(\xi)}}g(\phi,\psi)
\end{aleq}%
Once again, if we take the difference of equations (\ref{eq:62}.a) and (\ref{eq:62}.b), we deduce
the relation (\ref{eq:37c}). In order to simplify the equation (\ref{eq:63}) and to solve it, we
choose $\zeta_3(\xi)=0$. This is equivalent, considering (\ref{eq:60}), to requiring
$-a_1^{-1}\zeta_2(\xi)\zeta'_1(\xi)+\zeta'_2(\xi)=0$, which is satisfied if we choose
\begin{eqe}\label{eq:65}
\zeta_2(\xi)=\w_3 \exp(a_1^{-1}\zeta_1(\xi)).
\end{eqe}%
So the equation (\ref{eq:63}) transforms to
\begin{aleq}\label{eq:66}
&g_{\phi\psi}+\frac{c_1\pa{g_\phi+g_\psi}}{4a_2
\sin(2J(\phi,\psi))}-\Bigg(a_1\sin(2J(\phi,\psi))+\frac{1}{2a_2}(a_2^2-a_1^2)\cos(2J(\phi,\psi))\\
&+\frac{c_1a_1}{\sin(2J)\pa{-\la \sin(2J)+\mu\cos(2J(\phi,\psi))}}\Bigg)\pa{g_\phi-g_\psi}=0.
\end{aleq}%
We propose the solution of (\ref{eq:66}) in the form $g=g(J(\phi,\psi),K(\phi,\psi))$ with $J$
given by (\ref{eq:37c}) and $K=(2a_2)^{-1}c_1(\psi+\phi)$, which allows us to find the solution
\begin{eqe}\label{eq:67}
g(\phi,\psi)=\w_3\cos(2J(\phi,\psi))+\w_4\sin(2J(\phi,\psi))-\frac{c_1\w_4(\phi+\psi)}{2a_2}+\w_5.
\end{eqe}%
The solution expressed in term of $x$ and $y$ is
\begin{eqe}\label{eq:68}
g(x,y)=\pa{\w_3-\frac{\w_4 a_1}{a_2}}\cos(2J(\xi))-\frac{\w_4(xc_1+a_1)}{a_2}+\w_5.
\end{eqe}%
Since the compatibility condition (\ref{eq:66}) is satisfied, the system (\ref{eq:59}) can be
integrated after the substitution of $g(x,y)$ given by (\ref{eq:68}), $\zeta_2(\xi)$ by
(\ref{eq:65}) and $\zeta_3(\xi)=0$. The solution for $f(x,y)$ is given by
\begin{eqe}\label{eq:69}
f(x,y)=e^{\zeta_1(\xi)}\pa{-\frac{\w_3(-a_2\w_3+\w_4a_1)\cos(2J(\xi))}{a_1}-\frac{c_1\w_3^2y}{a_1}+\w_6},
\end{eqe}%
where $\zeta_1(\xi)$ is an arbitrary function of one variable and $\xi=a_1 x+a_2 y$. Finally, the
solution for the system (\ref{eq:1}.c), (\ref{eq:1}.d), is obtained by introducing (\ref{eq:58}),
(\ref{eq:68}) and (\ref{eq:69}) in (\ref{eq:44}). Redefining properly the parameters $\w_i$,
$i=3,4,5,6$, the solutions are
\begin{aleq}\label{eq:70}
&u(x,y)=c_4\pa{-a_2\cos(2J(\xi))+c_5y}+c_6,\\
&v(x,y)=c_4\pa{a_1\cos(2J(\xi))+(-c_5+c_1)x}+c_7,
\end{aleq}%
where the $c_i$, $i=1,\ldots,7$ are integration constants. So, we have obtained a solution of the
system (\ref{eq:1}) by defining the angle $\theta$ by (\ref{eq:formSolProp}), (\ref{eq:11}), the
pressure $\sigma$ by (\ref{eq:12}) and the velocities $u$ and $v$ by (\ref{eq:70}), with
$\xi(x,y)=a_1x+a_2y$. This solution includes the solution (\ref{eq:57}) to which it corresponds if
we choose $c_5=c_1$ in (\ref{eq:70}). A extrusion die corresponding to the solution (\ref{eq:70})
is shown in figure \ref{fig:3}. The parameter values that have been used are $k=0.8$, $c_1=2$,
$c_3=1$, $a_1=-10$, $a_2=10$, $c_4=1$, $c_5=1$, $c_6=10$, $c_7=-10$. To demonstrate how the
plasticity region changes when we vary the feeding velocities, we plotted two curves $C_1$ and
$C_2$ to delimit the plasticity region at the mouth of the tool. The curve $C_1$ corresponds to a
feeding velocity $U_0=3.54$ and the curve $C_2$ to a feeding velocity $U_0=3.84$, while the curve
$C_3$ is the plasticity limit at the output and the extraction velocity is $U_1=16.12$. This kind
of tool could be used to thin a plate for some ideal plastic material.
%
\subsection{Similarity solution for the angle $\theta$ and corresponding pressure
$\sigma$}\label{sec:3:2} In this section, we find solutions of the system (\ref{eq:1}) for which
the angle $\theta$ is a similarity solution. We propose the solution for $\theta$ in the form
provided by (\ref{eq:formSolProp}), but with the symmetry variable in the form
\begin{eqe}\label{eq:formXiDil}
\xi(x,y)=y/x.
\end{eqe}%
The solutions for $\theta$ and $\sigma$, which are obtained by this Ansatz, include for appropriate
parameters choices the invariant solution corresponding to the subalgebras $\mathcal{L}_{1,j}$, $j$
ranging from 5 to 10 and $L_{2,k}$ for $k$ ranging from 13 to 27. The introduction of
(\ref{eq:formSolProp}), with $\xi$ defined by (\ref{eq:formXiDil}), in the system (\ref{eq:1}.a),
(\ref{eq:1}.b), leads to the system
\begin{aleq}\label{eq:71}
&\s_x(x,y)=2(k/x) \pa{-\xi(x,y)
\cos(J(\xi(x,y)))+\sin(\xi(x,y))} J'(\xi(x,y)),\\
&\s_y(x,y)=-2(k/x)\pa{\xi(x,y)\sin(J(\xi(x,y)))+\cos(J(\xi(x,y)))}J'(\xi(x,y)).
\end{aleq}%
Considering the compatibility condition on mixed derivatives of $\sigma$ relative to $x$ and $y$,
we deduce from (\ref{eq:71}) the following ODE for the function $J$:
\begin{aleq}\label{eq:72}
&\pa{(\xi^2-1)\sin(2J(\xi))+2\xi\cos(2J(\xi))}J''(\xi)+2\pa{\xi\sin(2J(\xi))+\cos(J(\xi))}J'(\xi)\\
&+2\pa{-2\sin(2J(\xi))+(\xi^2-1)\cos(2J(\xi))}(J'(\xi))^2=0,
\end{aleq}%
which has the first integral
\begin{eqe}\label{eq:73}
\pa{(\xi^2-1)\sin(2J(\xi))+2\xi\cos(2J(\xi))}J'(\xi)=c_1,
\end{eqe}%
where $c_1$ is an integration constant. There are two cases to consider to solve the equation
(\ref{eq:73}).
\paragraph*{i}If $c_1\neq 0$ the solution of (\ref{eq:73}) is given
in implicit form by
\begin{eqe}\label{eq:74}
\frac{\pa{\tan(J(\xi))-\xi}\sqrt{c_1^2-1}}{\pa{\tan(J(\xi))\xi+1}c_1-\xi+\tan(J(\xi))}-\tan\pa{\frac{\sqrt{c_1^2-1}(c_2-J(\xi))}{c_1}}=0,
\end{eqe}%
where $c_2$ is an integration constant. The solution for $\sigma$ is obtained by integrating the
system (\ref{eq:71}) and taking into account the first integral (\ref{eq:73}). We find
\begin{eqe}\label{eq:75}
\s(x,y)=k\pa{\xi(x,y)\cos(2J(\xi(x,y)))-\sin(2J(\xi(x,y)))-2c_1\ln(x)}+c_3,\quad c_3\in \RR
\end{eqe}%
with $\xi(x,y)$ defined by (\ref{eq:formXiDil}) and $J(\xi)$ a solution of (\ref{eq:74}). So,
$\theta(x,y)=J(y/x)$ and $\s(x,y)$ are given by (\ref{eq:75}) and are solutions of the system
(\ref{eq:1}.a), (\ref{eq:1}.b), if the function $J(\xi)$ verify the algebraic equation
(\ref{eq:74}).
\paragraph*{ii}If the constant $c_1=0$ in the equation (\ref{eq:74}),
then the solution of (\ref{eq:72}) for $J$ is
\begin{eqe}\label{eq:76}
J(\xi)=-\frac{1}{2}\arctan\pa{\frac{2\xi}{\xi^2-1}}.
\end{eqe}%
We subsequently solve (\ref{eq:71}) for $\s(x,y)$ considering (\ref{eq:76}), which leads to the
solution of the system (\ref{eq:1}.a), (\ref{eq:1}.b), given by
\begin{eqe}\label{eq:77}
\theta(x,y)=\frac{1}{2}\arctan\pa{\frac{2xy}{x^2-y^2}},\qquad
\s(x,y)=-2k\arctan\pa{\frac{y}{x}}+c_2.
\end{eqe}%
\subsubsection{Additive separation for the velocities when
$c_1\neq0$.}\label{sec:3:2:1} Already knowing the solution $\theta(x,y)$, $\s(x,y)$, in the case of
$c_1\neq 0$, we still have to compute the solution for $u$ and $v$. A way to proceed is to suppose
that the solution is in the additive separated form
\begin{eqe}\label{eq:78}
u(x,y)=f(x,y)+F(\xi(x,y)),\qquad v(x,y)=g(x,y)+G(\xi(x,y)),
\end{eqe}%
where $\xi(x,y)=y/x$. We introduce (\ref{eq:78}) into the system (\ref{eq:1}.c), (\ref{eq:1}.d),
which gives
\begin{aleq}\label{eq:79}
&\pa{\sin(2J(\xi))-\cos(2J(\xi))}\xi F'(\xi)-\pa{\cos(2J(\xi))+\sin(2J(\xi))}\xi
G'(\xi)\\
&+\pa{\pa{f_y+g_x}\sin(2J(\xi))+\pa{ f_x-g_y}\cos(2J(\xi))}x=0,
\end{aleq}%
\begin{eqe}\label{eq:80}
\pa{f_x+g_y}x+G'(\xi)-\xi F'(\xi)=0,
\end{eqe}%
We must determine which functions $f$ and $g$ will reduce the equations (\ref{eq:79}),
(\ref{eq:80}), to a system of ODE for the one variable functions $F(\xi)$ and $G(\xi)$. To reach
this goal, we first use as annihilator the infinitesimal generator $D_1$ defined by (\ref{eq:2})
that we apply to the equations (\ref{eq:79}), (\ref{eq:80}), to eliminate the presence of the
functions $F$ and $G$. Indeed, $D_1$ annihilates any function of $\xi=y/x$. So, we obtain as
differential consequences some conditions on the functions $f$ and $g$. We can assume
$f_x(x,y)+g_y(x,y)\neq0$, otherwise we can show that the only possible solution is the trivial
constant solution for $u$ and $v$. Under this hypothesis, the previous conditions read
\begin{eqe}\label{eq:82}
f_x=-g_y+\zeta_1 (\xi)x^{-1},
\end{eqe}%
\begin{eqe}\label{eq:83}
f_y=-g_x+g_y\zeta_2(\xi)+\zeta_3(\xi)x^{-1},
\end{eqe}%
\begin{eqe}\label{eq:84}
\pa{g_y+D_1\pa{g_y}}\pa{\zeta_2(\xi)\sin(2J(\xi))-2\cos(2J(\xi))}=0,
\end{eqe}%
where the functions of one variable $\zeta_i$, $i=1,2,3$ are arbitrary. The left member of
(\ref{eq:84}) being composed of two factors, we must consider two possibilities.
\paragraph*{a.}We first suppose that
\begin{eqe}\label{eq:85}
g_y+D_{1}\pa{g_y}=0.
\end{eqe}%
In this case, we find that the functions $f$ and $g$ take the form
\begin{aleq}\label{eq:86}
f(x,y)=&-\int^{\xi(x,y)}\frac{\zeta_1(\xi)-\zeta_4'(\xi)}{\xi}d\xi+\omega_4\ln(y)-\omega_1y+\omega_5,\\
g(x,y)=&\zeta_4(\xi)+\omega_2\ln(x)+\omega_1 x+\omega_3,
\end{aleq}%
where the functions $\zeta_1(\xi), \zeta_4(\xi)$ are arbitrary and the functions $\zeta_2(\xi)$,
$\zeta_3(\xi)$, were chosen to solve the compatibility conditions on mixed derivatives of $f$
relative to $x$ and $y$. We now introduce the solution (\ref{eq:86}) in the system (\ref{eq:82}),
(\ref{eq:83}), which leads to an ODE system for $F$ and $G$, that we omit due to its complexity,
and for which the solutions take the form of quadratures
\begin{aleq}\label{eq:87}
F(\xi)=&\int
\pa{-\zeta'_4(\xi)+\frac{\zeta_1(\xi)}{\xi}+\frac{(\w_4+\w_2\xi)\sin(2J(\xi))}{\xi\pa{(\xi^2-1)\sin(2J(\xi))+2\xi\cos(2J(\xi))}}}d\xi+c_5,\\
G(\xi)=&\int\pa{-\zeta_1(\xi)+\xi F'(\xi)}d\xi+c_4,\qquad c_4,c_5\in \RR.
\end{aleq}%
The last step is to introduce (\ref{eq:86}) and (\ref{eq:87}) in the Ansatz (\ref{eq:78}). Then the
velocities are
\begin{aleq}\label{eq:88}
u(x,y)=&-\frac{1}{2}\frac{c_5\cos(2J(y/x))}{c_1}+\int^{y/x}\frac{c_6\sin(2J(\xi))J'(\xi)}{c_1\xi}d\xi+c_6\ln(y)-c_4
y+c_7,\\
v(x,y)=&c_5\ln(x)+c_4 x+\int^{y/x}c_1^{-1}
c_5\xi\sin(2J(\xi))J'(\xi)d\xi-\frac{1}{2}\frac{c_6\cos(2J(\xi))}{c_1}+c_8f,
\end{aleq}%
where the $c_i$ are integration constants. So, we obtain a solution of the system (\ref{eq:1}) by
defining the angle $\theta$ by (\ref{eq:formSolProp}), (\ref{eq:74}), the mean pressure $\sigma$ by
(\ref{eq:75}) and the velocities $u$ and $v$ by (\ref{eq:88}), with $\xi(x,y)=y/x$.
\paragraph*{b.}Suppose now that the condition (\ref{eq:84}) is satisfied by requiring
\begin{eqe}\label{eq:89}
\zeta_2(\xi)=2\cot(2J(\xi)).
\end{eqe}%
Then, applying the compatibility condition on mixed derivative of $f$ relative to $x$ and $y$ to
the equations (\ref{eq:82}), (\ref{eq:83}), and considering $\zeta_2$ given by (\ref{eq:89}), we
conclude that the function $g$ must solve the equation
\begin{aleq}\label{eq:90}
&g_{xx}(x,y)+2\cot(2J(\xi))g_{xy}(x,y)-g_{yy}(x,y)+4
x^{-1}\xi\pa{\xi+\cot^2(2J(\xi))}J'(\xi)g_y(x,y)\\
&+x^{-2}\pa{\zeta'_3(\xi)\xi+\zeta'_1(\xi)+\zeta_3(\xi)}=0.
\end{aleq}%
It's a hyperbolic equation everywhere in the domain where $J$ is defined. So, we introduce the
change of variable
\begin{aleq}\label{eq:91}
\phi(x,y)=&x\exp\pa{\int^{\xi(x,y)}\frac{\sin(2J(\xi))}{1+\cos(2J(\xi))+\xi\sin(2J(\xi))}d\xi},\\
\psi(x,y)=&x\exp\pa{\int^{\xi(x,y)}\frac{\sin(2J(\xi))}{-1+\cos(2J(\xi))+\xi\sin(2J(\xi))}d\xi},
\end{aleq}%
which brings the equation (\ref{eq:90}) to the simplified form
\begin{aleq}\label{eq:92}
&g_{\phi,\psi}+\frac{c_1}{2}\pa{\frac{\sin(2J(\phi,\psi))g_\phi}{\psi
(\cos(2J(\phi,\psi)))+1}-\frac{\sin(2J(\phi,\psi))g_\psi}{\phi(\cos(2J(\phi,\psi))-1)}}\\
&-\frac{1}{4}\frac{\sin(2J(\phi,\psi))\xi^2-\sin(2J(\phi,\psi))+2\xi\cos(2J(\phi,\psi))\pa{\zeta_3(\xi)+\zeta'_1(\xi)+\xi\zeta'_3(\xi)}}{\phi,\psi}=0,
\end{aleq}%
where $J(\phi,\psi)$ is defined by (\ref{eq:37c}). To solve the equation (\ref{eq:92}) more easily,
we define the function $\zeta_3$ by
\begin{eqe}\label{eq:93}
\zeta_3(\xi)=\frac{-\zeta_1(\xi)+J(\xi)+\omega_1}{\xi}.
\end{eqe}%
So, the solution of (\ref{eq:92}) is
\begin{eqe}\label{eq:94}
g(\phi,\psi)=-(1/2)\pa{\w_1-(1/2)\ln(\psi/\phi)}\cos\pa{c_1\ln(\psi/\psi)}-(1/4)c_1^{-1}\sin\pa{c_1\ln(\psi/\phi)},
\end{eqe}%
which, returning to the initial variables, takes the form
\begin{eqe}\label{eq:95}
g(x,y)=-(1/2)\pa{\w_1-(1/2)c_1^{-1}J(y/x)}\cos(2J(y/x))-(1/4)c_1^{-1}\sin(2J(y/x))+\w_2.
\end{eqe}%
After the introduction of the solution (\ref{eq:95}) for $g$, the function $f$ is given by
quadrature from the equations (\ref{eq:82}), (\ref{eq:83}). The obtained solution for $f$ is
\begin{eqe}\label{eq:96}
f(x,y)=\pa{\int^{\xi(x,y)}\frac{\pa{c_1\w_1-J(\xi)}\sin(2J(\xi))J'(\xi)}{c_1\xi}d\xi}-\int^{\xi(x,y)}\frac{\zeta_1(\xi)}{\xi}d\xi+(c_1+1)\w_1\ln(y)+\w_3.
\end{eqe}%
We now introduce (\ref{eq:95}), (\ref{eq:96}), in (\ref{eq:79}), (\ref{eq:80}), and get $F$ and $G$
by quadrature in the form
\begin{aleq}\label{eq:97}
F(\xi)&=\int^{\xi(x,y)}\frac{\zeta_1(\xi)}{\xi}d\xi+\int^{\xi(x,y)}\frac{(\w_1+J(\xi))\sin(2J(\xi))J'(\xi)}{c_1\xi}d\xi,\\
G(\xi)&=-(1/2)c_1^{-1}\pa{(\w_1+J(\xi))\cos(2J(\xi))+2\sin(2J(\xi))}.
\end{aleq}%
Finally, the substitution of (\ref{eq:95}), (\ref{eq:96}) and (\ref{eq:97}) in (\ref{eq:78})
provides the solution to (\ref{eq:1}.c), (\ref{eq:1}.d):
\begin{aleq}\label{eq:98}
u(x,y)=&(c_1+1)c_4\ln(y)+\int^{y/x}\frac{c_4(c_1+1)\sin(2J(\xi))J'(\xi)}{c_1\xi}d\xi+c_5,\\
v(x,y)=&-\frac{c_4(c_1+1)\cos(2J(y/x))}{c_1}+c_6,
\end{aleq}%
where the $c_i$ are integration constants. So, we have a solution of the system (\ref{eq:1}) by
implicitly defining the angle $\theta$ by (\ref{eq:formSolProp}), (\ref{eq:74}), the mean pressure
by $\sigma$ by (\ref{eq:75}) and the velocities $u$ and $v$ by (\ref{eq:98}), with $\xi(x,y)=y/x$.
\subsubsection{Additive separation for the velocities $u$ and $v$ when
$c_1=0$.}\label{sec:3:2:2}%
\begin{figure}[h]
\begin{center}
\includegraphics[width=8.5cm]{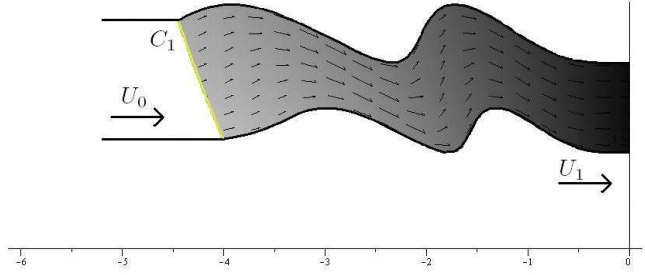}
\end{center}
\caption{Extrusion die corresponding to the solution (\ref{eq:12}), (\ref{eq:101}).}%
\label{fig:4}
\end{figure}

\begin{figure}[h]
\begin{center}
\includegraphics[width=8.5cm]{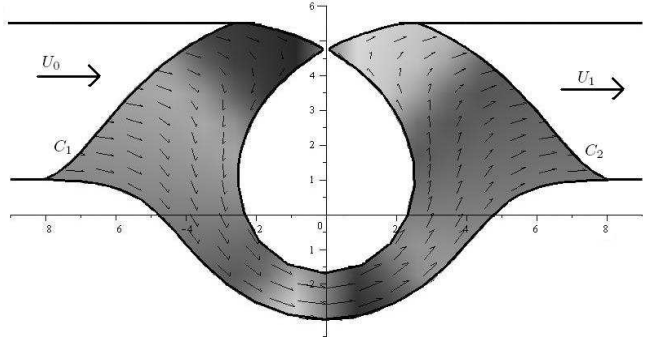}
\end{center}
\caption{Extrusion die corresponding to the solution (\ref{eq:12}), (\ref{eq:109}).}%
\label{fig:5}
\end{figure}

Now, we consider the case where $c_1=0$ in (\ref{eq:73}). Then the solutions for $\theta$ and
$\sigma$ are given by (\ref{eq:77}). We still suppose the solution for $u$ and $v$ in the form
(\ref{eq:78}). The procedure is the same as the previous case until we obtain the conditions
(\ref{eq:82}), (\ref{eq:83}) and (\ref{eq:84}). We must again consider two distinct cases.
\paragraph*{(a.)}We first suppose that the condition (\ref{eq:85}) is satisfied. Then, the functions $f$ and $g$
are defined by
\begin{aleq}\label{eq:99}
f(x,y)=&\int^{\xi(x,y)}(\zeta_2(\xi)+\xi)\zeta'_4(\xi)-\zeta_3(\xi)d\xi-\w_1y+\omega_3,\\
g(x,y)=&\zeta_4(\xi(x,y))+\w_1x+\w_2,\qquad \w_i\in\RR, \ i=1,2,3,
\end{aleq}%
where the $\zeta_i$, $=1,2,3$, are arbitrary functions of one variable and to simplify the
expression for $f$ and $g$ we have chosen
$\zeta_1(\xi)=\pa{1-\zeta_2(\xi)\xi-\xi^2}\zeta_4(\xi)-\xi\zeta_3(\xi)$. We substitute
(\ref{eq:99}) in the equations (\ref{eq:79}) and (\ref{eq:80}) to determine $F$ and $G$. We
conclude that $F(\xi)$ is an arbitrary function, while $G(\xi)$ is expressed in term of a
quadrature by
\begin{eqe}\label{eq:100}
G(\xi)=\int\pa{F'(\xi)\xi+\pa{\xi\zeta_2(\xi)-1+\xi^2}\zeta'_4-\xi\zeta_3(\xi)}d\xi.
\end{eqe}%
We finally obtain the solution $u$ and $v$ by introducing (\ref{eq:99}), (\ref{eq:100}) in
(\ref{eq:78}) and, to simplify, by choosing
$$\zeta_3(\xi)=-\pa{\zeta_2(\xi)+\xi}\zeta'_4(\xi),$$%
which gives
\begin{aleq}\label{eq:101}
u(x,y)=&-c_2 y+F'(y/x),\\
v(x,y)=&c_2 x+(y/x)F'(y/x)-F(y/x),
\end{aleq}%
where $F$ is a arbitrary function of one variable. A solution of the system (\ref{eq:1}) consists
of the angle $\theta$ and the pressure $\sigma$ defined by (\ref{eq:77}) with the velocities
defined by (\ref{eq:101}). For example, if we choose the arbitrary function to be an elliptic
function, that is
$$F(\xi)=\operatorname{dn}\pa{b_1 \pi \pa{1-e^{-b_2 \xi^2}},\varrho},$$%
and we set the parameters as $k=0.027$, $\epsilon=1$, $\omega_1=0$, $b_1=2$, $b_2=2$,
$\varrho=1/2$, then we can trace (see figure \ref{fig:4}) an extrusion die for a feeding speed of
$U_0=0.75$ and an extraction speed $U_1=1$. The curve $C_1$ on the figure \ref{fig:4} delimits the
plasticity region at the mouth of the tool, while the $x$-axis does the same for the output of the
tool. This type of tool could be used to undulate a plate. We can shape the tool by varying the
parameters. For example, we can vary the wave frequency setting the parameter $b_1$. Moreover, one
should note that if the modulus $\varrho$ of the elliptic function is such that $0\leq
\varrho^2\leq 1$, then the solution has one purely real and one purely imaginary period. For a real
argument $\chi$, we have the relations
$$\sqrt{1-\varrho^2}\leq dn(\chi,\varrho)\leq 1.$$%
\paragraph*{b.}Suppose that $\zeta_2(\xi)$ is defined by (\ref{eq:89}) and for simplification we
choose in particular
\begin{eqe}\label{eq:102}
\zeta_3(\xi)=\frac{\zeta_1(\xi)}{\xi}.
\end{eqe}%
Applying the mixed derivatives compatibility condition of $f$ to the equations (\ref{eq:82}),
(\ref{eq:83}), we get the following ODE for the function $g$:
\begin{eqe}\label{eq:103}
g_{xx}-g_{yy}-2\cot(2J(\xi))g_{xy}-\frac{4\xi J'(\xi)g_y}{x\sin^2(2J(\xi))}=0.
\end{eqe}%
By the change of variable
\begin{eqe}\label{eq:104}
\xi(x,y)=y/x,\qquad \eta(x,y)=x^2+y^2,
\end{eqe}%
we reduce the PDE (\ref{eq:103}) in term of $x$ and $y$, to the much simpler PDE in term of $\xi$
and $\eta$,
\begin{eqe}\label{eq:105}
g_{\xi \eta}+\frac{\xi g_\eta}{\xi^2+1}=0,
\end{eqe}%
which has the solution
\begin{eqe}\label{eq:106}
g(\xi,\eta)=\zeta_4(\xi)+\frac{\zeta_5(\eta)}{\sqrt{\xi^2+1}},
\end{eqe}%
where $\zeta_4$ and $\zeta_5$ are arbitrary functions of one variable. Then, we find the solution
for $f$ by integration of the PDE (\ref{eq:82}), (\ref{eq:83}), with $\zeta_3$ given by
(\ref{eq:102}),
\begin{eqe}\label{eq:107}
f(x,y)=-\int^{\xi(x,y)}\frac{\zeta_1(\xi)-\zeta'_4(\xi)}{\xi}d\xi-\frac{y\zeta_5(\eta(x,y))}{\sqrt{\eta(x,y)}}+c_2.
\end{eqe}%
By the substitution of (\ref{eq:106}) and (\ref{eq:107}) in the equations (\ref{eq:79}),
(\ref{eq:80}), we find that $F$ is an arbitrary function of one variable and $G$ is defined by
\begin{eqe}\label{eq:108}
G(\xi(x,y))=\int^{\xi(x,y)}\pa{-\zeta_1(\xi)+F'(\xi)\xi}d\xi+c_3.
\end{eqe}%
So, we introduce (\ref{eq:106}), (\ref{eq:107}) and (\ref{eq:108}) in (\ref{eq:78}) and after an
appropriate redefining of $\zeta_1$, $\zeta_4$ and $\zeta_5$, the solution to (\ref{eq:1}.c),
(\ref{eq:1}.d) is provided by
\begin{eqe}\label{eq:109}
u(x,y)=K'(y/x)-y H(x^2+y^2)+c_2,\qquad v(x,y)=-K'(y/x)+\xi K(\xi)+x H(x^2+y^2)+c_1,
\end{eqe}%
where $H$, $K$ are arbitrary functions of one variable. The velocities (\ref{eq:109}) together with
the angle and pressure defined by (\ref{eq:77}) solve the system (\ref{eq:1}). For example, a tool
corresponding to the solution (\ref{eq:109}) with $H(\eta)=2\exp(-0.1\eta)$, $K(\xi)=\xi$ and for
feeding and extraction speed given respectively by $U_0=1.05$ and $U_1=1.05$. It is shown in figure
\ref{fig:5}. The plasticity region limits correspond to the curves $C_1$ and $C_2$. This tool is
symmetric under the reflection $x\mapsto -x$. Moreover, the top wall of the tool almost makes a
complete loop, and this lets one suppose that we could make a ring in a material by extrusion.
\subsubsection{Multiplicative separation for the velocities $u$ and $v$ when $c_1\neq 0$.}
Consider the solutions of (\ref{eq:1}.a), (\ref{eq:1}.b), given by the angle $\theta(x,y)=J(y/x)$
with $J$ defined by (\ref{eq:74}) and the pressure $\sigma$ defined by (\ref{eq:75}). We require
the solutions for $u$ and $v$ in the multiplicative separated form
\begin{eqe}\label{eq:ms:1}
u(x,y)=f(x,y)F(\xi(x,y)),\quad v(x,y)=g(x,y)G(\xi(x,y)),
\end{eqe}%
where $\xi(x,y)=y/x$ and $f$, $g$, $F$, $G$ are to be determined. The Ansatz (\ref{eq:ms:1}) on the
velocities brings the system (\ref{eq:1}.c), (\ref{eq:1}.d) , to the form
\begin{aleq}\label{eq:ms:2}
&\br{f_yF+x^{-1}fF'(\xi)+g_xG(\xi)-x^{-1}\xi gG'(\xi)}\sin(2J(\xi))\\
&+\br{f_xF(\xi)-x^{-1}\xi
fF'-g_yG-x^{-1}gG'}\cos(2J(\xi))=0,\\
&\br{f_xF(\xi)+g_yG(\xi)}-\xi fF'(\xi)+gG'(\xi)=0.
\end{aleq}%
To reduce the PDE system (\ref{eq:ms:2}) to an ODE system involving $F$ and $G$ in term of $\xi$,
we act with the operator $D_1$, defined by (\ref{eq:2}), and annihilate the function of $\xi$
present in (\ref{eq:ms:2}). This leads to conditions on $f$ and $g$ that do not involve $F$, $G$
and their derivatives. There are three cases to consider, that is
\begin{aleq}\label{eq:ms:3}
&(a)\quad D_1\pa{g_y/g}\neq0,\qquad D_1\pa{\frac{D_1(f/g)}{D_1(xg_y/g)}}\neq0,\\
&(b)\quad D_1\pa{xg_y/g}\neq0,\qquad D_1\pa{\frac{D_1(f/g)}{D_1(xg_y/g)}}=0,\\
&(c)\quad D_1(x g_y/g)=0.
\end{aleq}%
In this paper, we present the details for the cases (a) and (c).
\paragraph*{(a)}We first suppose that the condition (\ref{eq:ms:3}.a) is satisfied. In this case,
the function $f$ must be solution of the PDE system
\begin{aleq}\label{eq:ms:4}
f_x=&-x^{-1}\xi\zeta_1(\xi)f+\zeta_2(\xi)g_y-x^{-1}\pa{\zeta_2(\xi)\zeta_1'(\xi)-\zeta_2'(\xi)}g,\\
f_y=&x^{-1}\zeta_1(\xi)f+\zeta_2(\xi)g_x-2\zeta_2(\xi)\cot(2J(\xi))g_y\\
&+\pa{\zeta_2(\xi)\zeta_1'(\xi)-\zeta_2'(\xi)}\frac{\xi\sin(2J(\xi))+2\cos(2J(\xi))}{x\sin(2J(\xi))}g,
\end{aleq}%
where $\zeta_1$, $\zeta_2$ are two arbitrary functions of one variable. For the system
(\ref{eq:ms:4}) to be compatible, the function $g$ must satisfy the PDE
\begin{aleq}\label{eq:ms:5}
&g_{xx}-2\cot(2J(\xi))g_{xy}-g_{yy}+2x^{-1}\zeta_2(\xi)^{-1}\pa{\xi+\cot(2J(\xi))}\zeta_3(\xi)g_x\\
&+2x^{-1}\br{\pa{\xi-\cot(2J(\xi))}\zeta_3(\xi)-2\xi\sin^{-2}(J(\xi))J'(\xi)}g_y\\
&+x^{-2}\zeta_2(\xi)^{-1}\Big([4\xi
J'(\xi)\sin^{-2}(2J(\xi))+\pa{\xi^2-1+2\xi\cot(2J(\xi))}\zeta_1'(\xi)\\
&-2\xi-2\cot(2J(\xi))]\zeta_3(\xi)-\pa{\xi^2-1+2\xi\cot(2J(\xi))}\zeta_3'(\xi)\Big) g=0,
\end{aleq}%
where we used the notation $\zeta_3(\xi)=\zeta_2(\xi)\zeta_1'(\xi)-\zeta_2'(\xi)$ to shorten the
expression. The equation (\ref{eq:ms:5}) is difficult to solve for arbitrary $\zeta_1$, $\zeta_2$,
but if we make the particular choice
\begin{eqe}\label{eq:ms:6}
\zeta_2(\xi)=\omega_2\exp(\zeta_1(\xi)),\qquad \omega_2\in \RR,
\end{eqe}%
then the PDE (\ref{eq:ms:5}) reduces to
\begin{eqe}\label{eq:ms:7}
g_{xx}-2\cot(2J(\xi))g_{xy}-g_{yy}-4x^{-1}\xi J'(\xi)\sin^{-2}(J(\xi))g_y=0,
\end{eqe}%
which is solved by the function
\begin{eqe}\label{eq:ms:8}
g(x,y)=\omega_3 x+\omega_4\cos(2J(y/x))+\omega_5,\quad \omega_3,\omega_4,\omega_5\in \RR.
\end{eqe}%
With $g$ given by (\ref{eq:ms:8}) and $\zeta_2(\xi)$ by (\ref{eq:ms:6}), the system (\ref{eq:ms:4})
is compatible. Consequently,  $f$ is expressed in term of a quadrature. We find
\begin{aleq}\label{eq:ms:9}
f(x,y)=&2\exp(\zeta_1(y/x))\omega_2\omega_4\int^{y/x}\xi^{-1}\sin(2J(\xi))J'(\xi)d\xi\\
&+\exp(\zeta_1(y/x))\pa{\omega_2\omega_3+2\omega_2\omega_4c_1\ln(y)+\omega_6}
\end{aleq}%
With $f$ given by (\ref{eq:ms:9}) and $g$ by (\ref{eq:ms:8}), the solutions for $F$ and $G$ of the
system (\ref{eq:ms:2}) are
\begin{eqe}\label{eq:ms:10}
F(\xi)=\omega_1\exp(-\zeta_1(\xi)),\qquad G(\xi)=\omega_2.
\end{eqe}%
By introducing (\ref{eq:ms:8}), (\ref{eq:ms:9}), (\ref{eq:ms:10}) in (\ref{eq:ms:1}) and redefining
the free parameters $\omega_i$, $i=1,2,3,4$, the solutions of (\ref{eq:1}.c), (\ref{eq:1}.d), for
the velocities $u$ and $v$ is
\begin{aleq}\label{eq:ms:11}
&u(x,y)=2c_4\int^{y/x}\xi^{-1}\sin(2J(\xi))J'(\xi)d\xi+c_5y+2c_1c_4\ln(y)+c_6,\\
&v(x,y)=-c_5x-c_4\cos(2J(y/x))+c_7,
\end{aleq}%
where the $c_i$, $i=1,\ldots,7$, are integration constants and $J$ is defined by (\ref{eq:74}). So,
we have a solution of the system (\ref{eq:1}) composed of the angle $\theta$ in the form
(\ref{eq:formSolProp}) with $J$ given implicitly by (\ref{eq:74}) together with the pressure
$\sigma$ (\ref{eq:75}) and the velocities (\ref{eq:ms:11}).
\paragraph*{(c)} Suppose now that the condition(\ref{eq:ms:3}.c) is
satisfied. In this case, the solution $g$ takes the form
\begin{eqe}\label{eq:ms:12}
g(x,y)=h_1(x)\zeta_1(\xi(x,y))
\end{eqe}%
and the function $f$ must be solution of
\begin{aleq}\label{eq:ms:13}
f_x=&x^{-1}\xi\zeta_2'(\xi)f+x^{-1}\zeta_3(\xi)h_1(x),\\
f_y=&-2x^{-1}h_1(x)\zeta_3(\xi)\cot(2J(\xi))+e^{-\zeta_2(\xi)}\pa{\int^{y/x}e^{\zeta_2(\xi)}\zeta_3(\xi)
d\xi-\omega_2}h_1'(x)\\
&-x^{-1}\br{\zeta_2'(\xi)+\xi\zeta_3(\xi)}.
\end{aleq}%
We omit, due to its complexity, the expression of the compatibility condition on the mixed
derivative of $f$ relative to $x$ and $y$. Nevertheless, making the specific choice
\begin{eqe}\label{eq:ms:14}
h_1(x)=1,\qquad
\zeta_2(\xi)=\ln\pa{\frac{\sin(2J(\xi))}{(\xi^2-1)\sin(2J(\xi))+2\xi\cos(2J(\xi))}}-\ln(\zeta_3(\xi)),
\end{eqe}%
the system (\ref{eq:ms:13}) turns out to be compatible and the solution to (\ref{eq:ms:13}) is
\begin{aleq}\label{eq:ms:15}
f(x,y)=&\zeta_3(y/x)\bigg(c_1^{-1}\pa{1-(y/x)^2-2(y/x)\cot(2J(y/x))}\int^{y/x}
\xi^{-1}\sin(2J(\xi))J'(\xi) d\xi \\
&+\br{1-(y/x)^2-2(y/x)\cot(2J(y/x))} (\ln y-\omega_4) \bigg).
\end{aleq}%
The substitution of (\ref{eq:ms:15}), (\ref{eq:ms:12}) and (\ref{eq:ms:14}) in the system
(\ref{eq:ms:2}) results in an ODE system for $F$ and $G$, omitted due to its complexity, which has
the solution
\begin{aleq}\label{eq:ms:16}
&F(\xi)=\frac{\omega_1\sin(2J(\xi))}{\zeta_3(\xi)\pa{(\xi^2-1)\sin(2J(\xi))+2\xi\cos(J(\xi))}},\\
&G(\xi)=\zeta_1(\xi)^{-1}\pa{\omega_1\cos(2J(\xi))-\omega_2}.
\end{aleq}%
We finally, obtain a solution for the system (\ref{eq:1}.c), (\ref{eq:1}.d), by the substitution of
(\ref{eq:ms:15}) and (\ref{eq:ms:16}), with $h_1(x)$, $\zeta_2(\xi)$, defined by (\ref{eq:ms:14}),
in (\ref{eq:ms:1}). This leads to
\begin{aleq}\label{eq:ms:17}
&u(x,y)=-c_4\ln(y)-c_1^{-1}c_4\int^{y/x}\xi^{-1}\sin(2J(\xi))J'(\xi)d\xi+c_5,\\
&v(x,y)=(1/2)c_1^{-1}\omega_1\cos(2J(y/x))+c_6,\qquad c_1,c_4,c_5,c_6 \in \RR.
\end{aleq}%
So, the system (\ref{eq:1}) is solved by the angle $\theta$ in the form (\ref{eq:formSolProp}) with
$J$ implicitly defined by (\ref{eq:74}), together with the pressure $\sigma$ (\ref{eq:75}) and the
velocities (\ref{eq:ms:17}).
\subsubsection{Multiplicative separation for the velocities $u$ and $v$ when $c_1=0$.}
Consider now the case where $c_1=0$ in (\ref{eq:73}) so the solution of (\ref{eq:1}.a),
(\ref{eq:1}.b), for $\theta$ and $\sigma$ is (\ref{eq:77}). We suppose that the velocities $u$ and
$v$ are in the form (\ref{eq:ms:1}). Introducing this form for the velocities and $\theta$ defined
by (\ref{eq:77}) in the equations (\ref{eq:1}.c), (\ref{eq:1}.d), leads to the system
(\ref{eq:ms:2}) which reduces to a ODE system for $F$ and $G$ if the functions $f$ and $g$ satisfy
the condition (\ref{eq:ms:3}). The three different constraints (\ref{eq:ms:3}) must be considered
separately.
\paragraph*{(a)}In the first case, where we consider that the conditions (\ref{eq:ms:3}.a) are satisfied, the
functions $f$ and $g$ must verify the system (\ref{eq:ms:4}), (\ref{eq:ms:5}). Changing the
variables $(x,y)$ to new variables $(\xi, \eta)$ defined by
\begin{eqe}\label{eq:ms:18.a}
\xi(x,y)=y/x,\qquad \eta(x,y)=x^2+y^2,
\end{eqe}%
 and considering $\theta$ given by (\ref{eq:77}), the system
(\ref{eq:ms:4}),(\ref{eq:ms:5}), becomes
\begin{aleq}\label{eq:ms:18}
&f_{\xi}=-\zeta_1(\xi)f-\xi^{-1}\zeta_2g_\xi-\xi^{-1}\pa{\zeta_2'(\xi)-\zeta_1(\xi)\zeta_2(\xi)}g,\\
&f_\eta=\xi\zeta_2(\xi)g_\eta,
\end{aleq}%
\begin{eqe}\label{eq:ms:19}
\frac{g_{\xi\eta}}{g_\eta}+\frac{\zeta_2'(\xi)}{\zeta_2(\xi)}+\zeta_1(\xi)+\xi(\xi^2+1)^{-1}=0,
\end{eqe}%
where  $\zeta_1(\xi)$, $\zeta_2(\xi)$ are arbitrary functions of one variable. The solution of the
system (\ref{eq:ms:18}), (\ref{eq:ms:19}), for $f$ and $g$ as function of $\xi$ and $\eta$ is
\begin{aleq}\label{eq:ms:20}
f(\xi,\eta)=&\xi\zeta_2(\xi)\pa{K(\xi)+\zeta_2(\xi)^{-1}(1+\xi^2)^{-1/2}\pa{H(\eta)e^{-\int\zeta_1(\xi)d\xi}}}\\
&-e^{-\int\zeta_1(\xi)d\xi}\Bigg(\int\xi^{-1}(\xi^2+1)\big[\zeta_2(\xi)+\pa{\zeta_1(\xi)+\xi(\xi^2+1)^{-1}\zeta_2(\xi)}H(\xi)\\
&+\zeta_2(\xi)K'(\xi)\big]d\xi-\omega_1\Bigg),\\
g(\xi,\eta)=&K(\xi)+\pa{e^{\int\zeta_1(\xi)d\xi}\zeta_2(\xi)\sqrt{\xi^2+1}},
\end{aleq}%
where the functions $K(\xi)$ and $H(\eta)$ are arbitrary functions of one variable. Introducing $f$
and $g$ expressed in the initial variables $x$, $y$, by the substitution of (\ref{eq:ms:18.a}) in
(\ref{eq:ms:12}), the system (\ref{eq:ms:2}) is reduced to an ODE system for the functions $F$ and
$G$ in term of $\xi$ which have the solution
\begin{eqe}\label{eq:ms:21}
F(\xi)=\omega_1e^{\int\zeta_1(\xi)d\xi},\qquad G(\xi)=-\omega_1\zeta_2 (\xi)e^{\int
\zeta_1(\xi)d\xi}.
\end{eqe}%
Finally, redefining
$$\zeta_2(\xi)=-(\omega_1K(\xi))^{-1}\exp\pa{\int\zeta_1(\xi) d\xi}Q(\xi),\qquad H(\eta)=\omega_1^{-1}\sqrt{\eta}P(\eta)$$%
where  $Q(\xi)$, $P(\eta)$ are arbitrary functions, and doing the substitution of (\ref{eq:ms:20})
and (\ref{eq:ms:21}) in (\ref{eq:ms:1}), we obtain the solution of (\ref{eq:1}.c), (\ref{eq:1}.d),
for velocities $u$ and $v$ given by
\begin{aleq}\label{eq:ms:22}
&u(x,y)=y P(x^2+y^2)-x^{-1}y Q(y/x)\int^{y/x}\xi^{-1}\pa{(\xi^2+1)Q'(\xi)+\xi Q(\xi)} d\xi+c_2,\\
&v(x,y)=Q(y/x)-x P(x^2+y^2),
\end{aleq}%
This solution for velocities together with $\theta$ and $\sigma$ defined by (\ref{eq:77}) solves
the initial system (\ref{eq:1}).
\paragraph*{(b)}Consider now that the functions $f$ and $g$ satisfy
the constraint (\ref{eq:ms:3}.b). Then they take the form
\begin{eqe}\label{eq:ms:22:1}
f(x,y)=\zeta_1(y/x),\qquad g(x,y)=\zeta_2(y/x),
\end{eqe}%
where $\zeta_1$ and $\zeta_2$ are arbitrary functions of one variable. Since the functions $f$ and
$g$ depend only on the symmetry variable and the velocities $u$ and $v$ have the form
(\ref{eq:ms:1}), it is equivalent to consider that
\begin{eqe}\label{eq:ms:22:2}
u(x,y)=F(\xi(x,y)),\qquad v(x,y)=G(\xi(x,y)).
\end{eqe}%
If we suppose that $u$ and $v$ are in the form (\ref{eq:ms:22:2}), then the solution of
(\ref{eq:1}) consists of the angle $\theta$ and the pressure $\sigma$ given by (\ref{eq:77})
together with the velocities
\begin{eqe}\label{eq:ms:22:3}
u(x,y)=F(y/x),\qquad v(x,y)=\int^{y/x}\xi F'(\xi)d\xi + c_3,
\end{eqe}
where $F$ is an arbitrary function of one variable.
\paragraph*{(c)}The third case to consider is when $f$ and $g$ obey
the conditions (\ref{eq:ms:3}.c), so they take the form
\begin{eqe}\label{eq:ms:23}
f(x,y)=\omega_3x^{1+\omega_2}(y/x)^{(1+\omega_2)/2}(x^{-1}(x^2+y^2))^{\omega_2/4},\qquad
g(x,y)=x^{1+\omega_2}\zeta_1(y/x),
\end{eqe}%
where $\zeta_1(\xi)$ is an arbitrary function of one variable. Then we introduce (\ref{eq:ms:23})
in (\ref{eq:ms:2}) and solve for $F$ and $G$. The solution is
\begin{aleq}\label{eq:ms:24}
F(\xi)=\omega_1\zeta_1^{-1}(\xi)\xi^{(1-\omega_2)/2}(\xi^2+1)^{\omega_2/4},\qquad
G(\xi)=(\xi^2+1)^{\omega_2/2}.
\end{aleq}%
Finally, substitution of $f$, $g$, $F$, $G$, given by (\ref{eq:ms:23}) and (\ref{eq:ms:24}), in
(\ref{eq:ms:2}) gives, after redefining the parameters $\omega_i$ in a convenient way, the solution
for $u$ and $v$ of the equations (\ref{eq:1}.c), (\ref{eq:1}.d),
\begin{eqe}\label{eq:ms:25}
u(x,y)=c_3y(x^2+y^2)^{\omega_2/2},\qquad v(x,y)=-c_4x(x^2+y^2)^{\omega_2/2},
\end{eqe}%
where $c_3$, $c_4$ are integration constants. The velocities $u$ and $v$ together with the angle
$\theta$ and the pressure $\sigma$ given by (\ref{eq:77}) constitute a solution for the system
(\ref{eq:1}). This solution is just a subcase of the previous one corresponding to the condition
(\ref{eq:ms:3}.a) and the choice $Q(\xi)=0$, $P(\eta)=\eta^{\omega_2/2}$ in (\ref{eq:ms:22}).
\subsection{Solution for $\theta$ in terms of the invariant $\sigma(x,y)+a_1 x+a_2
y$.}\label{sec:3:3} The subalgebras $\mathcal{L}_{2,28}$ to $\mathcal{L}_{2,31}$ do not have
symmetry variables, that is an invariant depending on $x$ and $y$ only, but they all have an
invariant in the form $\sigma+a_1 x+a_2 y$ for appropriate values of $a_1,a_2$. So, we suggest
$\theta$ in the form
\begin{eqe}\label{eq:110}
\theta(x,y)=J(\tau(x,y)),
\end{eqe}%
where
\begin{eqe}\label{eq:111}
\tau(x,y)=\sigma(x,y)+a_1 x+a_2 y.
\end{eqe}%
The function  $\tau(x,y)$ is to be determined since it depends on one of the unknown quantities,
that is $\sigma(x,y)$. Consequently, finding the function $\tau$ also determines the pressure
$\sigma(x,y)$. We first introduce $\theta$ given by (\ref{eq:110}) in the system (\ref{eq:1}.a),
(\ref{eq:1}.b), which leads, after the elimination of $\sigma(x,y)$ with the use of (\ref{eq:111}),
to the system
\begin{aleq}\label{eq:112}
&\tau_x(x,y)=\frac{a_1+2 k \pa{a_1 \cos(2J(\tau(x,y)))+a_2\sin(2J(\tau(x,y)))} J'(\tau(x,y))}{1-4k^2 J'(\tau(x,y))},\\
&\tau_y(x,y)=\frac{a_2+2 k \pa{a_1 \sin(2J(\tau(x,y)))-a_2\cos(2J(\tau(x,y)))}
J'(\tau(x,y))}{1-4k^2 J'(\tau(x,y))},
\end{aleq}%
of the two PDE for the function $\tau(x,y)$ using the compatibility conditions on mixed derivatives
of $\tau$ relative to $x$ and $y$, we deduce the second order ODE for the function $J(\tau)$
\begin{aleq}\label{eq:113}
&J''(\tau)+\frac{4k\pa{a_1^2+a_2^2}}{(a_2^2-a_1^2)\sin(2J(\tau))-2a_1a_2\cos(2J(\tau))}(J'(\tau))^3\\
&+2\frac{2a_1a_2\sin(2J(\tau))+(a_2^2-a_1^2)\cos(2J(\tau))}{(a_2^2-a_1^2)\sin(2J(\tau))-2a_1a_2\cos(2J(\tau))}(J'(\tau))^2=0.
\end{aleq}%
The solution of the previous equation is
\begin{eqe}\label{eq:114}
J(\tau)=\frac{1}{2}\arctan\pa{\frac{\mu\tau-\la
\sqrt{\la^2+\mu^2-\tau^2}}{\la\tau+\mu\sqrt{\la^2+\mu^2-\tau^2}}},
\end{eqe}%
where
\begin{eqe}\label{eq:115}
\la=(c_1/2)\pa{a_2^2-a_1^2},\qquad \mu=k(a_1^2+a_2^2)(a_2^2-a_1^2)^{-1}-c_1a_1a_2,\qquad c_1\in\RR.
\end{eqe}%
The solution of (\ref{eq:112}) for $\tau$ takes the implicit form
\begin{eqe}\label{eq:116}
\kappa_1 \sin(2J(\tau+c_2))+\kappa_2 \cos(2J(\tau+c_2))+x+\kappa_3 y+c_3=0,\qquad c_2,c_3\in \RR,
\end{eqe}%
where
\begin{eqe}\label{eq:117}
\kappa_1=c_1a_2+\frac{2ka_1}{a_1^2-a_2^2},\quad\kappa_2=\frac{c_1^2(a_1^2-a_2^2)^2-4k^2}{2c_1a_1
(a_1^2-a_2^2)+4ka_2},\quad\kappa_3=\frac{c_1a_2(a_1^2-a_2^2)+2k^2a_1}{c_1a_1 (a_1^2-a_2^2)+2ka_2}.
\end{eqe}%
Then we introduce (\ref{eq:114}) in (\ref{eq:116}) and we solve for $\tau$ to obtain the solution
in the explicit form
\begin{eqe}\label{eq:118}
\tau(x,y)=-\frac{\mu \kappa_1+\la \kappa_2}{\kappa_1^2+\kappa_2^2}(x+\kappa_3 y+c_3)\pm
\frac{\sqrt{\pa{\kappa_1^2+\kappa_2^2-(x+\kappa_3
y+c_3)^2}(\la\kappa_1-\mu\kappa_2)^2}}{\kappa_1^2+\kappa_2^2},
\end{eqe}%
where $\kappa_1,\kappa_2,\kappa_3$ are given by (\ref{eq:117}) and $\la,\mu$ are defined by
(\ref{eq:115}). Finally, the solution to the PDE (\ref{eq:1}.a) and (\ref{eq:1}.b) is
\begin{eqe}\label{eq:119}
\theta(x,y)=J(\tau(x,y)),\qquad \sigma(x,y)=\tau(x,y)-(a_1x +a_2y),
\end{eqe}%
with $\tau(x,y)$ defined by (\ref{eq:118}) and $J$ by (\ref{eq:114}).
\subsubsection{Additive separation for the velocities $u$ and $v$.}
Finally, we search for a solution to the PDE (\ref{eq:1}.c), (\ref{eq:1}.d), in the additive
separated form
\begin{eqe}\label{eq:120}
u(x,y)=f(x,y)+F(\tau(x,y)),\qquad v(x,y)=g(x,y)+G(\tau(x,y)),
\end{eqe}%
where $\tau(x,y)$ is given by  (\ref{eq:118}). We substitute (\ref{eq:120}) in the system
(\ref{eq:1}.c), (\ref{eq:1}.d), to obtain
\begin{aleq}\label{eq:121}
&\pa{f_y+\tau_yF'(\tau)+g_x+\tau_xG'(\tau)}\sin(2J(\tau))+\pa{f_x+\tau_xF'(\tau)-g_y-\tau_yG'(\tau)}\cos(2J(\tau))=0,\\
&f_x+\tau_xF'(\tau)+g_y+\tau_yG'(\tau)=0.
\end{aleq}%
We want this system to reduce to a ODE system for $\tau$. This is the case if the functions $f$ and
$g$ satisfy some conditions. In order to find these conditions, we act on the system (\ref{eq:121})
with the annihilator of $\tau$ defined by
\begin{eqe}\label{eq:122}
L_{\tau}=-\tau_y\del_x+\tau_x\del_y.
\end{eqe}%
The result is that $f$ and $g$ must obey the PDE system
\begin{aleq}\label{eq:123}
&f_x=\zeta_1(\tau)-g_y,\\
&f_y=-g_x+2\cot(2J(\tau))g_y+\zeta_2(\tau),
\end{aleq}%
where $\zeta_1$, $\zeta_2$ are arbitrary functions of one variable. The compatibility condition on
mixed derivative of $f$ relative to $x$ and $y$ leads to the following PDE for $g$:
\begin{eqe}\label{eq:124}
g_{xx}-2
\cot(2J(\tau))g_{xy}-g_{yy}+\frac{4\tau_xJ'(\tau)}{\sin^2(2J(\tau))}g_y+\zeta_1'(\tau)\tau_y-\zeta_2'(\tau)\tau_x=0.
\end{eqe}%
Any solutions $f$ and $g$ of the system consisting of (\ref{eq:123}) and (\ref{eq:124}) reduce the
system (\ref{eq:121}) to an ODE system for $F$ and $G$ in term of $\tau$. The general solution of
(\ref{eq:124}) is hard to find, but we give as an illustration a particular solution for $g$. We
make the specific choice
\begin{eqe}\label{eq:125}
\zeta_1(\tau)=\omega_1,\qquad\zeta_2(\tau)=-\frac{2\omega_2\cos(2J(\tau))}{\sin(2J(\tau))}+\omega_3,
\end{eqe}%
where $\omega_1$, $\omega_2$ and $\omega_3$ are free real parameters. Considering $\zeta_1(\tau)$,
$\zeta_2(\tau)$ given by (\ref{eq:125}), the solution for $g(x,y)$ of the equation (\ref{eq:124})
is
\begin{eqe}\label{eq:126}
g(x,y)=\omega_4 x+\omega_2 y,\qquad \omega_4\in \RR.
\end{eqe}%
We introduce (\ref{eq:125}) and (\ref{eq:126}) in the system (\ref{eq:123}) and we solve it to find
\begin{eqe}\label{eq:127}
f(x,y)=(\omega_1-\omega_2)x+(-\omega_4+\omega_3)y+\omega_5,\qquad\omega_6\in\RR.
\end{eqe}%
Using (\ref{eq:126}) and (\ref{eq:127}), the solutions $F(\tau)$ and $G(\tau)$ of the system
(\ref{eq:121}) are
\begin{aleq}\label{eq:128}
&F(\tau)=\kappa_4 \sin(2J(\tau))+\kappa_5\cos(2J(\tau))+c_4, &G(\tau)=\kappa_6
\sin(2J(\tau))+\kappa_7\cos(2J(\tau))+c_5,
\end{aleq}%
where $c_4,c_5$ are integration constants and
\begin{aleq*}\label{eq:129}
&\kappa_4=(\omega_1-\omega_2)\pa{a_2c_1+\frac{2ka_1}{a_1^2a_2^2}},\qquad
&&\kappa_5=(1/2)(a_1\omega_1-a_2\omega_3)c_1+\frac{a_2\omega_1-a_1\omega_3}{a_1^2-a_2^2},\\
&\kappa_6=(a_1c_1-\frac{2ka_2}{a_1^2-a_2^2})\omega_2,
&&\kappa_7=(1/2)(a_1\omega_3-a_2\omega_1)c_1+\frac{a_2\omega_3-a_1\omega_1}{a_1^2-a_2^2}.
\end{aleq*}%
Finally, the solutions of (\ref{eq:1}.c), (\ref{eq:1}.d), for the velocities $u$ and $v$ are
\begin{aleq}\label{eq:130}
&u(x,y)=(\omega_1-\omega_2)x+(-\omega_4+\omega_3)y+\kappa_4
\sin(2J(\tau(x,y)))+\kappa_5\cos(2J(\tau(x,y)))+c_4,\\
&v(x,y)=\omega_4 x+\omega_2 y+ \kappa_6 \sin(2J(\tau(x,y)))+\kappa_7\cos(2J(\tau(x,y))+c_5,
\end{aleq}%
where the function $J$ is defined by (\ref{eq:114}). The velocities (\ref{eq:130}) together with
the angle $\theta$ and the pressure $\sigma$ defined by (\ref{eq:119}) in which $J$ and $\tau$ are
respectively defined by (\ref{eq:114}) and (\ref{eq:118}) solve the system (\ref{eq:1}).
\section{Concluding remarks and future outlook}\label{sec:4}
The main goal of this paper was to construct analytic solutions of the system (\ref{eq:1})
modelling the planar flow of a ideal plastic material and to use the flow described by these
solutions to deduce possible shapes of extrusion dies. The group theoretic language appears to be a
very useful tool to obtain these solutions in the analysis of their admissible flow and their
properties. The main benefit of using group analysis is that we can find several classes of
solutions from a totally algorithmic procedure without using additional constraints but proceeding
only with the considered PDE system. We made an analysis of the symmetries of the system
(\ref{eq:1}) modelling a planar flow of an ideal plastic material and we computed its infinitesimal
symmetry generators. Subsequently, using these symmetries, we applied the SRM to find $G$-invariant
solutions of this system and this led to several solution classes. Many types of solutions were
found. For example we got rational algebraic (\ref{eq:31}), trigonometric (\ref{eq:31}), inverse
trigonometric (\ref{eq:11}), implicit (\ref{eq:74}) and some solutions in term of one or two
arbitrary functions of one variable (see (\ref{eq:101}) and (\ref{eq:109})), which can be chosen to
be Jacobi elliptic functions. Some others obtained solutions are expressed in term of quadratures
(\ref{eq:43b}) that must be solved numerically. However, it is more simple to numerically solve
these quadratures than to completely integrate the reduced equations numerically.
\paragraph*{} The application of the method
discussed above yielded several families of tools that can be used for the extrusion of ideal
plastic materials. For each solution family one may draw extrusion dies corresponding to compatible
choices (with the flow lines of the solution for the velocities $u$ and $v$) for given angles and
feeding velocities. Since we are free to choose the parameters in the solutions and to select which
lines of flow to use, we consequently have a wide variety of tools for each class of solutions. One
should note that the similarity solutions corresponding to the additive separation of the
velocities is particularly interesting since it is expressed in term of arbitrary functions and
this enlarges the variety of possible tools. In this paper, we have drawn some examples of possible
extrusion dies. For example we traced a tool that could curve a material in the figure \ref{fig:1}
and one in the shape of a deformed cross admitting two mouths to feed the tool. A tool that could
be used to thin a plate is shown in figure \ref{fig:3} and one to wave a plate in figure
\ref{fig:4}. A particularly interesting tool from the applications standpoint is shown in figure
\ref{fig:5}. It might be used to shape a ring by extrusion.
\paragraph*{}The method of characteristics \cite{systQuasi} has already been used to
obtain some solutions of the system (\ref{eq:1}). However, with this method one is constrained to
require the curves limiting the plasticity region to be characteristic curves. This is not, in
general, the case for $G$-invariant solutions. Generally, when we use the method of characteristics
to find solutions of (\ref{eq:1}), we obtain $u$ and $v$ by numerical integration along the
characteristics (as in \cite{Czyz:1}), while the use of the SRM ables us to find some classes of
analytical solutions not covered by the method of characteristics. In this context, a natural
question arise: what physical insight does one gain from exact analytic particular solutions.  A
partial answer is that they show up qualitative features that might be very difficult to detect
numerically: the existence of different types of periodic solutions and different type of localized
solutions. Stable solutions should be observable and should also provide a good starting point for
perturbative calculations.
\paragraph*{}The next step of this work is a systematic Lie group
analysis of the symmetries of the nonstationary system in a $2+1$ dimensional modelling plane flow
of an ideal plastic material. We expect that the SRM will lead to wider classes of physically
important
solutions and consequently to new extrusion dies.%
\section*{Acknowledgments}The author is greatly indebted to professor A.M. Grundland (Universit\'e du
Qu\'ebec \`a Trois-Rivi\`eres and Centre de Recherche Math\'ematiques de l'Universit\'e de
Montr\'eal) for several valuable and interesting discussions on the topic of this work. This work
was supported by research grants from NSERC of Canada.
\begin{landscape}
\begin{table}[h]
\begin{center}
\caption{List of one-dimensional subalgebras and their invariants. To ease the notation, we noted
the invariants $F_1=F$, $F_2=G$, $F_3=H$. The index $i$ in  $\mathcal{L}_{i,j}$ corresponds to the
subalgebra dimension and the index $j$ to the number of the subalgebra.} \fontsize{8}{10}
\selectfont
\begin{tabular}{|l l l l l l l|}
  \hline\hline
  & &  & symmetry &\multicolumn{3}{c|}{Invariants}\\ \cline{5-7}
  No & subalgebra & parameters & variable $\xi$ & $F$ & $G$ & $H$ \\
  \hline
  \multirow{2}{*}{$\mathcal{L}_{1,1}$} & $\ac{\al P_1+a^\al (1-\dd_{0a})P_2+\bb P_3}$ & $a\neq -\al$ & $a^\al (\dd_{0a}-1)x +\al y$ & $u$ & $v$ & $\s- \frac{\bb(x+y)}{\al + a^\al
(1-\dd_{0a})}$ \\
   & where $\dd_{0a}=\left\{\begin{aligned} &1\quad &&\text{if }a=0\\ &0 &&\text{if }a\neq 0
\end{aligned}\right.$ & $a=-\al$ & $y$ & $u$ & $v$ & $\s-\bb x$ \\
\hline \multirow{2}{*}{$\mathcal{L}_{1,2}$}& \multirow{2}{*}{$\ac{\al P_1+a^\al P_2+\bb P_3+\al P_4+(1-\al) P_5}$} & $\al\neq -a$ & $a^\al x -\al y$ & $u-\al\frac{x+y}{\al+a^\al}$ & $v-\frac{(1-\al)(x+y)}{\al+a^\al}$ & $\s-\bb\frac{x+y}{\al +a^\al}$\\
& & $\al=1=-a$ & $x+y$ & $-x+u$ & $v$ &$\s-x$\\
  \hline \multirow{2}{*}{$\mathcal{L}_{1,3}$} & \multirow{2}{*}{$\ac{B+\al P_1+a^\al P_2+\bb P_3}$} & $\al\neq1\neq -a$ & $a^\al x -\al y$ & $u-\frac{a^{\al}(x^2-y^2)-2\al xy}{2\pa{a^{2\al}+\al}}$ & $v-\frac{\al (x^2-y^2)+2a^\al
xy}{a^{2\al}+\al}$ & $\s-\frac{\bb (x+y)}{\al+a^\al}$\\
  & & $\al=1=-a$ & $x+y$ & $x\pa{\frac{x}{2}+y}+u$ & $\frac{x^2}{2}+v$ & $\s-\bb x$\\
  \hline
  \multirow{2}{*}{$\mathcal{L}_{1,4}$} & \multirow{2}{*}{$\ac{D_2+\al P_1 +a^\al P_2+b P_3}$} & $a\neq -\al$ & $-a^\al x+\al y$ & $u e^{-\frac{x+y}{\al+a^\al}}$ & $ve^{-\frac{x+y}{\al+a^\al}}$ & $\s-\frac{x+y}{\al+a^\al}$\\
  &  & $$ & $x+y$ & $ue^{-x}$ & $ve^{-x}$ & $\s-bx$\\
  $\mathcal{L}_{1,5}$ & $\ac{D_1+a^\al D_2+b P_3}$ & & $\frac{y}{x}$ & $x^{-a}u$ & $x^{-a}v$ & $\s-b\ln x$\\
  $\mathcal{L}_{1,6}$ & $\ac{D_1+D_2+B+bP_3}$ & & $\frac{y}{x}$ & $\frac{u}{x}+\frac{y}{x}\ln x$ & $\frac{v}{x}-\ln x$ & $\s-b\ln x$\\
  $\mathcal{L}_{1,7}$ & $\ac{D_1+aP_3+\al P_4+b^\al P_5}$ & & $\frac{y}{x}$ & $u-\al \ln x$ & $v-b^\al \ln x$ & $\s-a\ln x$\\
  $\mathcal{L}_{1,8}$ &$\ac{\al P_4+a^\al P_5+\bb P_3}$ &  & $\xi_1=x,\ \xi_2=y$ & $(\al + a^\al)\s-\bb (u+v)$ & $\al v-a^\al u$ & $$\\
  $\mathcal{L}_{1,9}$ & $\ac{D_2+b P_3}$ & & $\xi_1=x,\ \xi_2=y$ & $\s-b\ln u$ & $\frac{u}{v}$ & $$\\
  $\mathcal{L}_{1,10}$ & $\ac{B+\al P_3}$ & & $\xi_1=x, \xi_2=y$ & $\frac{\al}{y}u+\s$ & $\frac{x}{y}u+v$ & $$\\
  \hline\hline
\end{tabular}
 \end{center}
 \label{tab:sadim1}
\end{table}
\end{landscape}
\begin{landscape}
\begin{table}
\begin{center}
\caption{List of two-dimensional subalgebras and their invariants. To ease the notation, we denoted
the invariants $F_1=F$, $F_2=G$, $F_3=H$. The index $i$ in  $\mathcal{L}_{i,j}$ corresponds to the
subalgebra dimension and the index $j$ to the number of the subalgebra.} \label{tab:sadim2a}
\fontsize{8}{8} \selectfont%
\begin{tabular}{|l l l l l l l|}
\hline\hline
  & &  & symmetry &\multicolumn{3}{c|}{Invariants}\\ \cline{5-7}
  No & subalgebra & parameters & variable $\xi$ & $F$ & $G$ & $H$ \\
  \hline
  $\mathcal{L}_{2,1}$ & $\ac{P_1+aP_2+\al P_4,P_3}$ &  & $a x +y$ & $-\al x+u$ & $v$ &  \\
  \hline
  \multirow{4}{*}{$\mathcal{L}_{2,2}$} & \multirow{4}{*}{$\ac{\al P_1+a^\al P_2+\bb P_3,P_3+\w P_4+b^\w P_5}$} & $\al+ a^\al\neq 0$ & $-a^\al x + \al y$ & $-b^\w u+\w v$ & $\s-\frac{\bb
(x+y)}{\al+a^\al}-\frac{\dd (u+v)}{\w+b^\w}$ &  \\
\cline{3-7}
   &  & $\begin{aligned}&\al\neq 1\neq -a,\\ &\w = -b=1\end{aligned}$ & $-a^\al x + \al y$ & $u+v$ & $\s-\frac{\bb (x+y)}{\al+a^\al}-\frac{u-v}{2}$ &  \\
 \cline{3-7}
   &  & $\begin{aligned}&\al= -a=1,\\ &\w +b^\w\neq 0\end{aligned}$ & $-a^\al x + \al y$ & $-b^\w u+\w v$ & $\s-\bb\frac{x-y}{2}-\frac{u+v}{\w+b^\w}$ &  \\
 \cline{3-7}
   &  & $\begin{aligned}&\al= -a=1,\\ &\w =-b=1\end{aligned}$ & $-a^\al x + \al y$ & $u+ v$ & $\s-\bb\frac{x-y}{2}-\frac{u-v}{2}$ &  \\
\hline
  \multirow{2}{*}{$\mathcal{L}_{2,3}$} & \multirow{2}{*}{$\ac{B+\al P_1+a^\al P_2,\bb P_4+b^\bb P_5+P_3}$} & $a^\al\neq 0$ & $-a^\al x + \al y$ & $\begin{aligned}&a^\al u + v - \al
\frac{x^2}{2}\\ &-(1-\al) xy+\frac{y^2}{2}-(\bb+b^\bb)\s\end{aligned}$ & $\begin{aligned}&a^\al u -v +\al \frac{x^2}{2}+(1-\al) xy\\ &+\frac{y^2}{2}-(\bb-b^\bb)\s\end{aligned}$ &  \\
\cline{3-7}
   &  & $\begin{aligned}&\al =1,\\ &a=0\end{aligned}$ & $y$ & $u + v - \frac{x^2}{2}+ xy+(\bb+b^\bb)\s$ & $u - v + \frac{x^2}{2}+
xy+(\bb-b^\bb)\s$ &  \\
\hline
   \multirow{2}{*}{$\mathcal{L}_{2,4}$} & \multirow{2}{*}{$\ac{D_2+\al P_1+a^\al P_2+b P_3,P_5}$} & $\al+ a^\al\neq 0$ & $-a^\al x + \al y$ & $ue^{-\frac{x+y}{\al+a^\al}}$ &  &$\s-\frac{b(x+y)}{\al+a^\al}$\\
\cline{3-7}
 & & $\al = -a =1$ & $x+y$ & $ue^{-\frac{x-y}{2}}$ & $\s-\frac{x-y}{2}$
 &\\
 \hline
  \multirow{2}{*}{$\mathcal{L}_{2,5}$} & \multirow{2}{*}{$\ac{\al P_1+a^\al P_2+\bb P_3,\w P_4+b^\w P_5}$} &  $\begin{aligned}&b\neq 0,\\ &\al+ a^\al\neq 0\end{aligned}$ & $-a^\al x + \al y$ & $ue^{-\frac{x+y}{\al+a^\al}}$ &  &$\s-\frac{b(x+y)}{\al+a^\al}$\\
\cline{3-7}
  & & $\begin{aligned}&b\neq 0,\\ &\al=-a=1\end{aligned}$ & $ x + y$ & $ue^{-\frac{x+y}{\al+a^\al}}$ & & $\s-\frac{b (x-y)}{2}$\\
  \hline
   $\mathcal{L}_{2,6}$ & $\ac{B+\al P_1+a^\al P_2,\bb P_4+b^\bb P_5}$ & $b\neq 0$ & $-a^\al x + \al y$ & $\begin{aligned}&2\pa{\bb v -b^\bb v}+\al\pa{(b^\bb a -\bb)x^2-2b^\bb xy}\\ &+(1-\al)\pa{-b^\bb y^2-2\bb xy}\end{aligned}$  &  & $\s$\\
   \hline
   \multirow{2}{*}{$\mathcal{L}_{2,7}$} & \multirow{2}{*}{$\ac{B+\al P_1+a^\al P_2+P_3,\bb P_4+b^\bb P_5}$} & $\begin{aligned}&b\neq 0,\\ &a^\al\neq 0\end{aligned}$ & $-a^\al x + \al y$ & $\bb v -b^\bb u-\al \bb \frac{x^2}{2}-b^\bb \frac{y^2}{2 a^\al}-(1-\al) \bb x y$ &   & $\s-\frac{y}{a^\al}$\\
 \cline{3-7}
    & & $\begin{aligned}&b\neq 0,\\ &\al=1,\ a=0\end{aligned}$ & $y$ & $\bb v -b^\bb u-\bb \frac{x^2}{2}-b^\bb xy$ &  & $\s-x$\\
    \hline
     \end{tabular}
     \end{center}
\end{table}
\begin{table*}
\fontsize{8}{8} \selectfont%
\begin{center}
\begin{tabular}{|l l l l l l l|}
\multicolumn{7}{c}{\phantom{$\qquad\mathcal{L}_{2,6}\qquad\qquad \ac{\al P_1+a^\al P_2+\bb P_3,P_3+\w P_4+b^\w P_5}\quad\al\neq 1\neq -a -b^\w u+\w v 2\pa{\bb v -b^\bb v}+\al\pa{(b^\bb a -\bb)x^2-2b^\bb xy} a^\al u -v +\al \frac{x^2}{2}+(1-\al) xy \s-\frac{b(x+y)}{\al+a^\al} \s-\frac{b(x+y)}{\al+a^\al}$ }}\\
\multicolumn{7}{|c|}{Table \ref{tab:sadim2a} continued}\\
\hline
     \multirow{2}{*}{$\mathcal{L}_{2,8}$} &  \multirow{2}{*}{$\ac{B+\al P_1+a^\al P_2+P_3,\bb P_4+b^\bb P_5},$} & $\begin{aligned}&b\neq 0,\\ &\al+a^\al\neq 0\end{aligned}$ & $-a^\al x + \al y$ &  & $v-\al \frac{x^2}{2}-(1-\al) x y$  & $\s-\frac{x+y}{\al+a^\al}$\\
     \cline{3-7}
      & & $\begin{aligned}&b\neq 0,\\ &\al=1,\ a=-1\end{aligned}$ & $x + y$ &  & $v- \frac{x^2}{2}$ & $\s-\frac{x-y}{2}$\\
    \hline
     $\mathcal{L}_{2,9}$ & $\ac{B+\al P_1+a^\al P_2+P_3,\bb P_4+b^\bb P_5}$ &  & $\begin{aligned}&b\neq 0\\ &-a^\al x + \al y\end{aligned}$ &  & $v-\al \frac{x^2}{2}-(1-\al) x y$ & $\s-\frac{x+y}{\al+a^\al}$\\
      \hline
     \multirow{2}{*}{$\mathcal{L}_{2,10}$} & $\ac{B+\al P_1+a^\al P_2+P_3,\bb P_4+b^\bb P_5},$ & $\begin{aligned}&b\neq 0,\\ &\al+a^\al\neq 0\end{aligned}$ & $-a^\al x + \al y$ &  & $v$ & $\s-\frac{x+y}{\al+a^\al}$\\
     \cline{3-7}
     & & $\begin{aligned}&b\neq 0,\\ &\al=1,\ a=-1\end{aligned}$ & $x + y$ &  & $v$ & $\s-\frac{x-y}{2}$\\
     \hline
     \multirow{2}{*}{$\mathcal{L}_{2,11}$} & $\ac{D_2+\al P_1+a^\al P_2+bP_3,P_4+c P_5}$ & $\al+a^\al\neq 0$ & $-a^\al x + \al y$ &  & $(v- c u)e^{\frac{x+y}{\al + a^\al}}$ &$\s-b\frac{x+y}{\al+a^\al}$\\
      \cline{3-7}
     & & $\al=1,\ -a=1$ & $x + y$ &  & $(v- c u)e^{\frac{x-y}{\al -a^\al}}$ & $\s-b\frac{x-y}{\al-a^\al}$\\
     \hline
     $\mathcal{L}_{2,12}$ & $\ac{D_1+D_2+bP_3,B+ P_3}$ &  & $\frac{y}{x}$ & $\s+\frac{u}{y}-b \ln y$ & $\s-\frac{v}{x}-b \ln x$ &\\
      \hline
     $\mathcal{L}_{2,13}$ & $\ac{D_1+D_2+bP_3,B+ P_3}$ &  & $\frac{y}{x}$ & $u-\al \ln x$ & $v-a^\al \ln x$ &\\
     \hline
     $\mathcal{L}_{2,14}$ & $\ac{D_1+D_2+bP_3,B+ P_3}$ &  & $\frac{y}{x}$ & $u- \ln x$ & $b v+a \ln x-\s$ &\\
      \hline
     $\mathcal{L}_{2,15}$ & $\ac{D_1+D_2+bP_3,B+ P_3}$ &  & $\frac{y}{x}$ & $u-\ve \s+ \ve a \ln x$ & $v-b u -\ln x$ &\\
     \hline
     $\mathcal{L}_{2,16}$ & $\ac{D_1+a D_2+bP_3,P_3}$ &  & $\frac{y}{x}$ & $ux^{-a}$ & $vx^{-a}$ &\\
     \hline
     $\mathcal{L}_{2,17}$ & $\ac{D_1 a P_3,D_2+b P_3}$ &  & $\frac{y}{x}$ & $b \ln u+a \ln x -\s$ & $\frac{v}{u}$ &\\
     \hline
     $\mathcal{L}_{2,18}$ & $\ac{D_1+ D_2+B,P_3}$ &  & $\frac{y}{x}$ & $\frac{u+y \ln x}{x}$ & $\frac{v+x \ln x}{x}$  & $$\\
      \hline
     $\mathcal{L}_{2,19}$ & $\ac{D_1+ a P_3,D_2}$ &  & $\frac{y}{x}$ & $\frac{v}{u}$ &  &  $\s-a \ln x$\\
      \hline
     $\mathcal{L}_{2,20}$ & $\ac{D_1+a D_2 +b P_3,P_4+c P_5}$ & $c\neq 0$ & $\frac{y}{x}$ & $(c u - v) x^{-a}$ & &  $\s-b \ln x$\\
     \hline
     $\mathcal{L}_{2,21}$ & $\ac{D_1+ D_2+B+a P_3,P_4+b P_5}$ & $b\neq 0$ & $\frac{y}{x}$ & $\frac{b u - v+(x+b y)\ln x}{x}$ &  &  $\s-a \ln x$\\
     \hline
     $\mathcal{L}_{2,22}$ & $\ac{D_1+a D_2+b P_3,B}$ &  & $\frac{y}{x}$ & $\pa{u+\frac{x}{y}u}x^{-a}$ &  &  $\s-b \ln x$\\
     \hline
     $\mathcal{L}_{2,23}$ & $\ac{D_1+\al P_4 + a^{\al}P_5+b P_3,B}$ &  & $\frac{y}{x}$ & $\frac{-(\al x + a^\al y)\ln x+x u+ y v}{y}$ &  &  $\s-b \ln x$\\
     \hline
     $\mathcal{L}_{2,24}$ & $\ac{D_1+ P_4 + aP_3,P_5}$ &  & $\frac{y}{x}$ & $u-a\ln x$ &  &  $\s-b \ln x$\\
     \hline
     $\mathcal{L}_{2,25}$ & $\ac{D_1+ P_5 ,P_4+a P_5}$ &  & $\frac{y}{x}$ & $v-b u-\ln x$ &  &  $\s-a \ln x$\\
     \hline
     $\mathcal{L}_{2,26}$ & $\ac{D_1+ a D_2+b P_4 ,P_4}$ &  & $\frac{y}{x}$ &  & $v x^{-a}$ &  $\s-b \ln x$\\
     \hline
     $\mathcal{L}_{2,27}$ & $\ac{D_1+D_2+B+a P_3 ,P_4}$ &  & $\frac{y}{x}$ &  & $\frac{v}{x}-\ln x$ &  $\s-a \ln x$\\
  \hline\hline
  \end{tabular}
  \end{center}
\end{table*}
\end{landscape}
\begin{landscape}
\begin{table}
\begin{center}
\caption{List of one-dimensionnal subalgebras and their invariants. To ease the notation, we
denoted the invariants $F_1=F$, $F_2=G$, $F_3=H$. The index $i$ in  $\mathcal{L}_{i,j}$ corresponds
to the subalgebra dimension and the index $j$ to the number of the subalgebra.}
 \label{tab:sadim2b}
\begin{tabular}{|l l l l l l l|}
  \hline\hline
  & &  & symmetry &\multicolumn{3}{c|}{Invariants}\\ \cline{5-7}
  No & subalgebra & parameters & variable $\xi$ & $F$ & $G$ & $H$ \\
  \hline
     $\mathcal{L}_{2,28}$ & $\ac{P_1+aP_2+b P_3 ,P_2+P_4+c P_5+\al P_3}$ &  & $(\al a -b)x-\al y+\s$ & $a x - y +u$ & $(a x-y)c+v$ &  \\
     \hline
     $\mathcal{L}_{2,29}$ & $\ac{P_1+aP_2+ P_3 ,P_1+b P_2+P_4+c P_5}$ &  & $\frac{x b-y}{a-b}+\s$ & $u+\frac{-a x +y}{a-b}$ & $v+\frac{c (-a x +y)}{a-b}$ &  \\
     \hline
     $\mathcal{L}_{2,30}$ & $\ac{P_1+aP_3 ,P_2+b P_3}$ & $a^2+\al\neq 0$ & $\s-(a x - \al y)$ & $u$ & $v$ &  \\
     \hline
     $\mathcal{L}_{2,31}$ & $\begin{aligned}&\left\{D_2+\al P_1+(1-\al)P_2+aP_3,\right. \\ &\quad\left.(1-\al)P_1+b^{1-\al}P_2+P_3\right\}\end{aligned}$ &  & $\begin{aligned}&\s-\al(ax+y)\\ &-(1-\al)\\ &\cdot\pa{(1-ab^{1-\al})x +\al y}\end{aligned}$ & $u e^{\pa{-(1-\al)(-b^{1-\al}x+y)-\al x}}$ & $v e^{\pa{-(1-\al)(-b^{1-\al}x+y)-\al x}}$ &  \\
     \hline
     $\mathcal{L}_{2,32}$ & $\ac{D_1\al P_4+a^\al P_5+ b P_3,\bb P_1+c^{\bb}P_2}$ & $b\neq 0$ & $\s- a \ln (x b^\bb-\bb y)$ & $\frac{u-x}{x^\bb-\bb y}$ & $\frac{v-\bb b x+(\bb-1)y}{x b^\bb-\bb y}$ &  \\
     \hline
     $\mathcal{L}_{2,33}$ & $\ac{D_1+D_2+b P_3,\bb P_1+c^{\bb}P_2}$ & $b\neq 0$ & $\s- b \ln (x c^\bb-\bb y)$ & $u-\al\ln(c^\bb x-\bb y)$ & $v-a^\al\ln(c^\bb x-\bb y)$  &  \\
     \hline
     $\mathcal{L}_{2,34}$ & $\ac{D_1+a D_2+bP_3,\bb P_1+b^{\bb}P_2}$ & $b\neq 0$ & $\s- b \ln (c^\bb x-\bb y)$ & $u(c^\bb x - \bb y)^{-a}$ & $v(c^\bb x - \bb y)^{-a}$ &  \\
  \hline\hline
\end{tabular}
\end{center}
\end{table}
\begin{table}
\fontsize{8}{8} \selectfont%
\begin{center}
\caption{Subalgebra sort by the form of the invariant solution. The functions $f,g,\phi,\g$ depend
on $x,y$ and $F,G$ depend on $\xi$.}\label{tab:ansatzs}
\begin{tabular}{|c|c|c|c|c|c|c|c|c|c|c|}
\hline\hline & & \multicolumn{8}{|c|}{Ansatzs}\\
\hline
 No. & $\xi$ & $\begin{aligned} &u=f+F,\\ &v=g+G\end{aligned}$ & $\begin{aligned}&u=fF,\\ &v=gG\end{aligned}$ & $\begin{aligned}&u=f+\phi F,\\ &v=g+\g G\end{aligned}$ & $u=h(v)+fF$ & $u=G h(v)+fF$ & $u=h(v)+f+\phi F$ & $v=q(u)+g+\g G$ & $v=gq(u)+G$\\
  \hline
1 &  $a_1 x+ a_2 y$ & $\begin{aligned}&\mathcal{L}_{1,1}, \mathcal{L}_{1,2},\\ &\mathcal{L}_{1,3}, \mathcal{L}_{2,1}\\ &\mathcal{L}_{2,2},\mathcal{L}_{2,3}\end{aligned}$  & $\begin{aligned}&\mathcal{L}_{1,4}, \\ \end{aligned}$ &  & $\mathcal{L}_{2,4},\mathcal{L}_{2,5}$  & $\mathcal{L}_{2,6},\mathcal{L}_{2,7}$ & & $\begin{aligned}&\mathcal{L}_{2,8},\mathcal{L}_{2,9}\\ &\mathcal{L}_{2,10},\mathcal{L}_{2,11}\end{aligned}$  & \\
  \hline
 2&  $\frac{y}{x}$ & $\begin{aligned}&\mathcal{L}_{1,7},\mathcal{L}_{2,13},\\ &\mathcal{L}_{2,14},\mathcal{L}_{2,15}\end{aligned}$ & $\begin{aligned}&\mathcal{L}_{1,5},\mathcal{L}_{2,16} \\ &\mathcal{L}_{2,17}\end{aligned}$ & $\begin{aligned}&\mathcal{L}_{1,6}, \mathcal{L}_{2,12},\\ &\mathcal{L}_{2,18}\end{aligned}$  & $\begin{aligned}&\mathcal{L}_{1,9},\mathcal{L}_{2,19},\\ &\mathcal{L}_{2,20} \end{aligned}$ & $\mathcal{L}_{2,22}, \mathcal{L}_{2,23}$ & $\begin{aligned}&\mathcal{L}_{1,8},\mathcal{L}_{2,21},\mathcal{L}_{2,24}\\ \end{aligned}$ & $\begin{aligned}& \mathcal{L}_{2,25},\mathcal{L}_{2,26}\\ &\mathcal{L}_{2,27}\end{aligned}$ &$\mathcal{L}_{1,10}$\\
  \hline
  3 & $a_1 x + a_2 y + \s$  & $\begin{aligned}&\mathcal{L}_{2,28}, \mathcal{L}_{2,29}\\ &\mathcal{L}_{2,30}\end{aligned}$ & $\mathcal{L}_{2,30}$ &  &  &  &  &   & \\
  \hline
  4 & $a_3 \ln(a_1 x+a_2 y)+\s$ & $\mathcal{L}_{2,33}$ & $\mathcal{L}_{2,34}$ & $\mathcal{L}_{2,32}$ &  &  &   &  & \\
  \hline\hline
\end{tabular}
\end{center}
\end{table}
\end{landscape}
%
%
\bibliographystyle{unsrt}

\end{document}